\documentclass[aps,pre,twocolumn,nobibnotes,superscriptaddress,longbibliography]{revtex4-2}

\usepackage{amssymb}
\usepackage{graphicx}
\usepackage{amsmath,amsfonts}
\usepackage{hyperref}
\usepackage[utf8] {inputenc}
\usepackage{xcolor}  
\usepackage[normalem]{ulem}
\usepackage[T1]{fontenc} 
\usepackage{multirow}
\usepackage{booktabs}
\usepackage{threeparttable}
\usepackage{tabularx}
\usepackage{float}
\hypersetup{
	colorlinks=true,
	allcolors=blue
}
\begin{document}
	\title{Recovery of activation propagation and self-sustained oscillation abilities in stroke brain networks }
	
	\date{\today}
	\author{Yingpeng Liu}
	\affiliation{School of Physics and Electronic Engineering, Jiangsu University, Zhenjiang, Jiangsu, 212013, China}
	
	\author{Jiao Wu}
	\affiliation{School of Mathematical Sciences, Jiangsu University, Zhenjiang, Jiangsu, 212013, China}
	
	\author{Kesheng Xu}
	\affiliation{School of Physics and Electronic Engineering, Jiangsu University, Zhenjiang, Jiangsu, 212013, China}
	
	\author{Muhua Zheng}
	\email[]{zhengmuhua163@gmail.com}
	\affiliation{School of Physics and Electronic Engineering, Jiangsu University, Zhenjiang, Jiangsu, 212013, China}
	
\begin{abstract}
Healthy brain networks usually show highly efficient information communication and self-sustained oscillation abilities. However, how the brain network structure affects these dynamics after an injury (stroke) is not very clear. The recovery of structure and dynamics of stroke brain networks over time is still not known precisely. Based on the analysis of a large number of strokes' brain network data, we show that stroke changes the network properties in connection weights, average degree, clustering, community, etc. Yet, they will recover gradually over time to some extent. We then adopt a simplified
reaction-diffusion model to investigate stroke patients' activation propagation and self-sustained oscillation abilities. Our results reveal that the stroke slows the adoption time across different brain scales, indicating a weakened brain's activation propagation ability. In addition, we show that the lifetime of self-sustained oscillatory patterns at three months post-stroke patients' brains significantly departs from the healthy one. Finally, we examine the properties of core networks of self-sustained oscillatory patterns, in which the directed edges denote the main pathways of activation propagation. Our results demonstrate that the lifetime and recovery of self-sustaining patterns are related to the properties of core networks, and the properties in the post-stroke greatly vary from those in the healthy group. Most importantly, the strokes' activation propagation and self-sustained oscillation abilities significantly improve at one year post-stroke, driven by structural connection repair. This work may help us to understand the relationship between structure and function in brain disorders.
\end{abstract}
\maketitle

\section{Introduction}\label{intro}
Comprehending the relationship between human brain architecture and function is one of the central focuses in network neuroscience nowadays. In the last decades, extensive investigation of the architecture of the human connectome~\cite{Bullmore2009,Rubinov2010,Fornito2016} has demonstrated typical characteristics of complex networks, including the small-world effect~\cite{Watts1998,Sporns2004,Hagmann2007}, heterogeneous degree distributions~\cite{Gong2008,Crossley2014}, high clustering coefficient~\cite{Hagmann2007},  rich club phenomenon~\cite{VandenHeuvel2011}, modular structure~\cite{Sporns2016,Meunier2010}, and so on. 
These intricate structural
features of the brain networks support the functions of highly efficient neuronal information communication among different brain regions~\cite{avena2018communication,bullmore2012economy,zheng2020geometric}.

Evidence shows that the performance of various cognitive tasks is closely related to the dynamics of activation propagation and the emergence of self-sustained
oscillations in healthy brains~\cite{tononi2010information,sporns2013network,hilgetag2020hierarchy,meunier2010modular}. In terms of activation propagation, Mi\v{s}i\'{c} \textit{et al.}~\cite{mivsic2015cooperative} investigate the global spreading dynamics on anatomical brain networks and show how the structure of the network shapes the transmission of activation. They demonstrated that the early spreading was promoted by hub regions and a backbone of pathways, while the spread
of cascades was accelerated by the shortest path. Competing cascades evolved integrated by joining in polysensory associative regions. Therefore, understanding the complicated network structure is significant in comprehending highly efficient neuronal information spreading behaviors in the brain.

On the other hand, the dynamic patterns of activation propagation often exhibit the phenomena of self-sustained
oscillations in healthy brains~\cite{carr1993processing,buonomano2009state,mongillo2008synaptic,mi2013long,xu2014simplified,xu2013controlling,roxin2004self}, in which the activated nodes will propagate the firing signal to their neighbors via the structural connections and eventually emerge distinct activity patterns~\cite{bansal2019cognitive,honey2009predicting,kinouchi2006optimal,brunel1999fast}. Barzon, \textit{et al.} have uncovered that the emergence of collective oscillations is driven by the criticality and network structure in a whole-brain stochastic model~\cite{barzon2022criticality}. Moreover, the self-sustained
oscillation patterns in resting-state brain networks are usually connected to the rhythms, and healthy brain networks possess the typical features of multiscaled rhythms. Different rhythms can frequently switch in the brain~\cite{huo2022time}. Thus, understanding the dynamical patterns is fundamental for comprehending the rhythm mechanism.  

Although much work has addressed the abilities of highly efficient activation propagation and self-sustained oscillations in healthy brains, how the brain network alters these dynamic behaviors after an injury such as a stroke remains poorly understood. In addition, the recovery of structure and dynamics after focal brain injuries (stroke) over time is still not known precisely. Neurological dysfunctions shall alter these dynamic behaviors if highly efficient activation propagation and self-sustained oscillations are fundamental properties of healthy brains. Researchers have discovered that the stroke's brain network has fewer inter-hemispheric connections, and loss of these ties is the dominant aberrant pattern in stroke~\cite{griffis2019structural}. Besides, some studies have focussed on the changes in structure and dynamics due to brain injuries, such as disrupted criticality with realistic connectomes of stroke patients~\cite{rocha2022recovery,rocha2018homeostatic}
, Alzheimer’s disease~\cite{jiang2018impaired}, and during epileptic seizures~\cite{meisel2012failure,meisel2020antiepileptic}. However, we know little about the influence of the stroke brain on activation propagation and self-sustained oscillation processes.

In this work, we study stroke-affected brains' structure, activation propagation, and dynamical patterns using the  dataset from Ref.~\cite{rocha2022recovery}. The brains' structural connectivity data was measured at $3$ months after stroke ($Pat. \ t_1$), $1$ year after stroke ($Pat. \ t_2$), and in healthy controls ($Con$). This data allows us to measure departures
from normal structural connectivity, activation propagation and self-sustained oscillation abilities, and the recovery of corresponding dynamics behaviors over time. We show that stroke changes the network properties in connection weights, average degree, average clustering coefficient, community, etc., leading to the significantly weakened brain's activation propagation efficiency and self-sustaining oscillation ability at three months post-stroke. All these aspects recover to some extent after twelve months, driven by brain network connections remodeling.    

The arrangement of this paper is as follows. In Section \ref{II}, we briefly introduce the structural network dataset of stroke-affected brains used in this study. In addition, we present an activation propagation model on the structural brain networks. In Section \ref{SC}, we conduct a statistical analysis of the structural network within the stroke dataset. In Section \ref{Adoption time}, we demonstrate that stroke-induced structural damage weakens the brain's activation propagation abilities. In Section \ref{TLSOP}, we discover that stroke reduces the main pathways of activation flow within the brain and diminishes the brain's rhythm-related self-sustaining oscillation ability. We also found that brain function can experience a certain degree of recovery with the structural connections restored. Finally, Section \ref{IV} provides the discussion and conclusion.


\section{Stroke Dataset and The activation propagation model}\label{II}

\subsection{Connectomes of post‑stroke brains}\label{dataset}

An extensive prospective longitudinal stroke investigation contributes to the imaging and behavioral data~\cite{siegel2016disruptions,corbetta2015common,ramsey2017behavioural}. We use the structural connectivity data from
$79$ patients in previous study~\cite{rocha2022recovery}, where 
the patients' connectomes were measured at $3$ months ($Pat. \ t_1$) and $1$ year ($Pat. \ t_2$) after stroke onset. Notice that although there are $79$ participants, only $34$ took part in both $t_1$ and $t_2$ measurements/scans. So, we have $54$ samples in the $t_1$ and $59$ in $t_2$. 
Besides, the study also contains $28$ healthy controls ($Con$), and 18 were measured twice at a 3-month interval. So, we have a total sample size of $46$ in healthy controls. 
As the variability and difference in structural connectivity between the two time points in healthy controls are very small~\cite{rocha2022recovery}, we combine the datasets measured twice (sample size $N_g=46$) to represent the control group average as a baseline. By doing so, we can increase the sample size of structural connectivity for the healthy group, which will minimize the influence of experimental fluctuations/errors
and provide a more robust statistical reference for comparisons with stroke patients.    

Each connectome can be regarded as a network, and the nodes represent the cortical regions of interest (ROIs). In each connectome, the entire brain is divided into $324$ regions (ROIs), i.e., the number of nodes or network size $N=324$. The weight of the link, ($\tilde{w}_{ij}$), represents the fiber
density, which is defined as the number of white-matter fiber tracts connecting ROIs $i$ and $j$ normalized by the product of their average surface and average fiber length~\cite{hagmann2008mapping}. Taking into account the homeostatic plasticity principles~\cite{deco2014local,hellyer2016local,abeysuriya2018biophysical}, the structural connectivity matrix from Ref.~\cite{rocha2022recovery} has been normalized for each individual with 
\begin{equation}\label{eq:1}
	w_{ij} = \frac{\tilde{w}_{ij}}{\sum_{j}{\tilde{w}_{ij}}}.
\end{equation}
Previous studies have shown that the above normalization will minimize the variability of the neural activity patterns and the critical point
of the stochastic model for distinct subjects~\cite{rocha2022recovery,rocha2018homeostatic}. In addition, it will promote the comparison of the statistical results for single subjects~\cite{rocha2022recovery} (see more details about the influence of normalized connectivity weights in Figs.~\ref{comment1} and \ref{comment2} in Appendix~\ref{B}). Therefore, we use the normalized structural connectivity matrix in this work.

\subsection{Description of the activation propagation model}\label{RDM}
\begin{figure*}[!ht]
	\centering
	\includegraphics[width=1.0\textwidth]{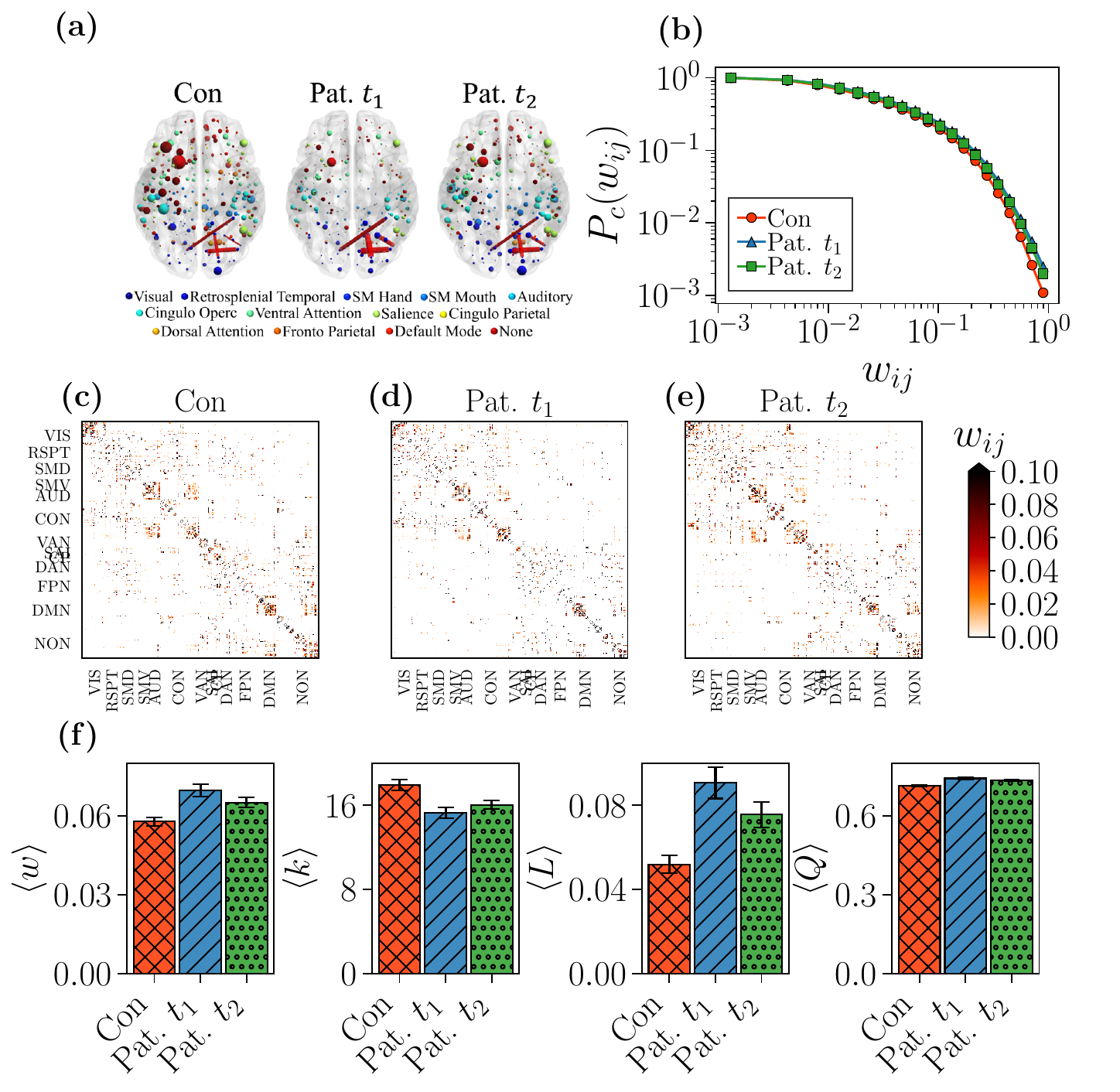}
	\caption{Network properties in the post-stroke brain. (a) From left to right, the axial view illustrations show the distribution of brain ROIs for three representative individuals from healthy controls ($Con$), stroke patients at 3 months post-stroke ($Pat. \ t_1$), and stroke patients at 1 year post-stroke ($Pat. \ t_2$). The colors indicate the brain regions and the size of the nodes is proportional to the degree of the nodes. Some edges between the same ROIs in the visual region are highlighted with the thickness of the edges proportional to the weights. (b) The complementary cumulative distribution functions of weights in $Con$, $ Pat. \ t_1 $ and $ Pat. \ t_2 $ group. (c-e) Structural connectivity matrices for the same representative individuals in (a). (f)  From left to right, we show the average weight $\langle w \rangle$, average degree $\langle k \rangle$, average path length $\langle L \rangle$,  and modularity $\langle Q \rangle$, respectively. Error bars represent each group's standard error of the mean (SEM). } 
	\label{figure1}
\end{figure*}

We adopt a simplified
reaction-diffusion (spreading) model on the empirical brain networks to study cortical activation propagation~\cite{huo2022time}. 
One advantage of studying the brain networks' spreading dynamics is that the simple model could provide a much-needed theoretical framework for analyzing how activation propagation processes unfold within the networks~\cite{avena2018communication}. 
For example, the spreading model may better reveal the organizational principles of brain networks shaping global communication and facilitating integrative function~\cite{mivsic2015cooperative}. Therefore, we use the spreading model to mimic the neural signal communications process. In the model,  each node in the model has an active (A) or inactive (I) state. The activation propagation process has two main steps: reaction and diffusion. In the reaction step, we focus only on the activated nodes, while the diffusion step concentrates on the nodes that have not yet been activated (i.e., the inactive state). To describe this process more precisely, we introduce the variable $ S_i(t) $ to represent the state of node $ i $ at time $ t $. Specifically, we assume that $ S_i(t)=1 $ when a node $i$ is in an activated state and $S_i(t)=0$ in an inactivated state. In the reaction step, a node with $ S_i(t)=1 $ has a probability $ p $ of transitioning from activated to inactivated and $ 1-p $ to stay activated. This transition reflects the dynamic characteristics of the system's nodes and lays the foundation for the subsequent diffusion process. Mathematically, this is expressed as:
\begin{equation}\label{eq:2}
	S_i(t+1) = \begin{cases} 
		0, & \text{with probability } p \\
		1, & \text{with probability } 1 - p 
	\end{cases}
\end{equation} 
for $ S_i(t) = 1 $. Without loss of generality, we set $p = 0.5 $ in our study.

In the diffusion step, if the total input coupling strength of an inactive node exceeds the activation threshold $ \omega_c $, the node will become activated. This process can be represented as follows:
\begin{equation}\label{eq:3}
	S_i(t+1) = \Theta \left( \sum_{j \in N_i} w_{ij} \cdot S_j(t) - \omega_c \right),
\end{equation}
for an inactivated node $ S_i(t) = 0 $. Here, $N_i$ is the  neighborhood of node $i$ and $ \Theta $ denotes the Heaviside function, which is $1$ if its argument is positive and $0$ otherwise.

We can easily acquire the mean-field version of the dynamics for the above activation propagation model. Let's define $ \rho_i(t) $ as the probability that node $ i $ is activated at time $t$. On the one hand, if a node is in the activated state, it remains activated with a probability $ 1-p $. Thus, the probability that node $i$ remains activated after a one-time step is $(1-p) \rho_i(t)$. On the other hand, the probability that an inactive node $ i $ becomes activated in the diffusion step depends on the sum of the connection weights of the activated neighbors $j$. Specifically, the node will be activated if the total input exceeds the threshold $ \omega_c $. So, the probability that node $ i $ becomes activated in the diffusion process is $(1-\rho_i(t)) \Theta\left(\sum_{j \in N_i} w_{ij} \rho_j(t) - \omega_c\right)$.  Thus, the mean-field theory equation is: 
\begin{equation}\label{eq:6}
	\rho_i(t\!+\!1) \!=\! (1\!-\!p)\rho_i(t)\!+\! (1\!-\!\rho_i(t))\Theta\left(\sum_{j \in N_i} w_{ij} \rho_j(t)\!-\! \omega_c\right).
\end{equation}
The analytical solution of Eq.~($\ref{eq:6}$) is difficult to obtain. However, by studying the diffusion step at the beginning with a small initial seed, we can obtain an approximate solution for the lower bound $\omega'_c$ (i.e.,  critical point) on a fully connected network with uniform connection weights. To maintain the diffusion process, one needs 
\begin{equation}\label{eq:condition}
	\sum_{j \in N_i} w_{ij} \rho_j(t)-\omega'_c \geq 0.		
\end{equation}
For a fully connected network with uniform connection weights, we have 
\begin{equation}\label{eq:condition2}
	\omega'_c \leq	N\langle w\rangle\rho_0,  	
\end{equation}
where $\rho_0$ is the initial activation ratio (i.e., initial seed) and $\langle w\rangle$ is the average weight. We have verified the theory by using simulations on synthetic networks. As shown in Fig.~\ref{figureS1} in Appendix
\ref{A}, the simulation results are consistent with the theory very well. From Eq.~($\ref{eq:condition2}$), we can see that the behavior of activation propagation depends on the seeds initially selected and the activation threshold. 
Notice that although our mathematical discussion in Eq.~($\ref{eq:condition2}$) does not apply directly to real brain networks, the results make us realize that the behavior of activation propagation depends on the seeds initially selected and the activation threshold. Therefore, in the next section, we will fix the seeds at the initial time step so that we can compare the statistical results in different groups better.

\section{Result}\label{III}

\begin{figure*}[!ht]
	\centering
	\includegraphics[width=1.0\textwidth]{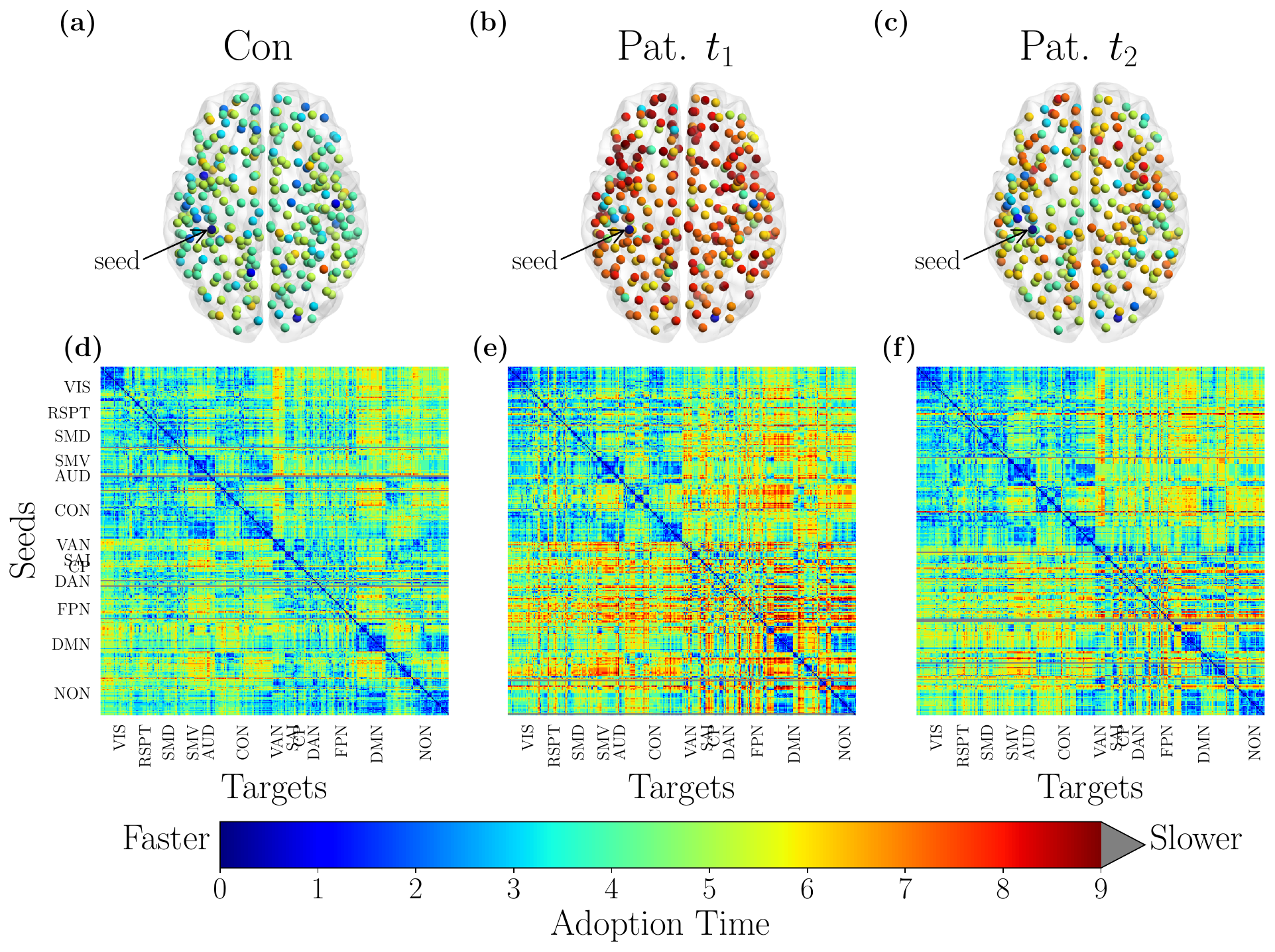}
	\caption{The spread patterns and corresponding adoption time matrices for representative individuals from different groups. (a-c) The spread patterns reflect the first adoption time of each node under a single seed perturbation (marked by the arrow) for three representative individuals. Blue nodes mean the adoption times are faster, while red ones are slower. (d-f)  Adoption time matrices for the same individual in (a-c) show the first time it carries for a perturbation started at a selected seed node (rows) to reach another one (columns). We have arranged the matrices with ROIs based on the cortical parcellation of Gordon \textit{et al}~\cite{gordon2016generation}. The results are obtained from $200$ independent realizations for each seed. Here $\omega_c = 0.03$ is fixed.}
	\label{figure2}
\end{figure*}

\begin{figure*}[ht]
	\centering
	\includegraphics[width=1.0\textwidth]{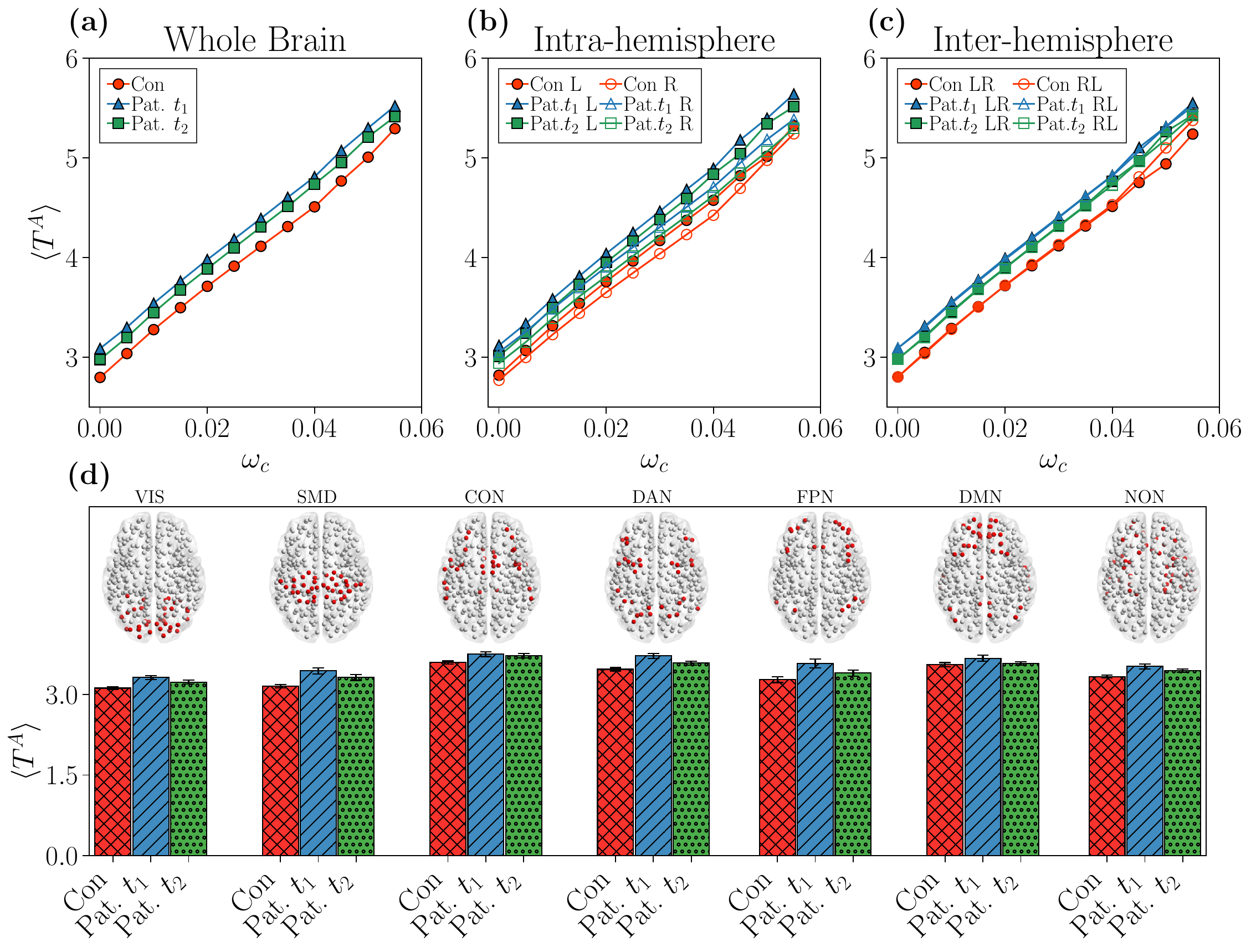}
	\caption{The average adoption time for different groups. (a-c) show the average adoption time as a function of threshold $\omega_c$ in the whole brain, intra-hemispheric, and inter-hemispheric regions, respectively. Error bars representing the corresponding SEM are smaller than the symbol size. (d) shows each group's average adoption time and corresponding SEM(bottom) at $\omega_c = 0.03$ with seven considered brain regions.  Nodes in the same cortical parcellation are highlighted in red color(top). }
	\label{figure3}
\end{figure*}
\subsection{Structural alterations and recovery of the post-stroke brain}\label{SC}



Lesions undoubtedly depart from normal structural connectivity, but the structural alterations and recovery over time after stroke are still not known precisely. Stroke typically begins with an acute disruption of blood flow to a specific brain region, leading to direct damage to the brain parenchyma~\cite{carmichael20163,carmichael2010targets}. Although the structural damage from stroke is localized, it induces both local and widespread alterations in brain function and structure~\cite{shi2019global,carter2012use}. Group analyses in stroke patients have shown that the majority of strokes impact subcortical/cortical structural connectivity, including both white and gray matter pathways~\cite{guggisberg2019brain,corbetta2015common}.

We here examine the brain network properties in the healthy control group ($Con$), the patient group after $3$ months post-stroke ($Pat. \ t_1$), and $1$ year post-stroke ($Pat. \ t_2$). In Fig.~\ref{figure1}(a), we show the axial view for three selected representative individuals from the three groups (Individual No.~$12$ in $Con$ and the same individual, No.~$12$, in $Pat. \ t_1$ and $Pat. \ t_2$). The brain network is divided into $13$ regions, each containing a different number of ROIs. Nodes belonging to the same region are assigned with the same color. The size of each node is proportional to its degree $k$. We can observe that for the same ROIs, the sizes of nodes in the $Con$ are generally larger than those in the $Pat. \ t_1$ and $Pat. \ t_2$. Additionally, the sizes of nodes for the same individual at $1$ year post-stroke are greater than the case at three months post-stroke, indicating the formation of new connections between specific ROIs after one year. 
Furthermore, we highlight several edges between the same ROIs in the visual region, in which the thickness of the edges is proportional to the weights. We find the weights' alterations and recovery by comparing the edges' thickness at $3$ months and $1$ year post-stroke. To make it clear, in Fig.~\ref{figure1}(b), we show the complementary cumulative distribution of the weights of edges, $P_c(w_{ij})$ in the $Con$, $ Pat. \ t_1 $ and $ Pat. \ t_2 $ group. The distribution is slightly lower in the healthy control group than in the stroke one. Interestingly, the distribution in $Pat. \ t_2$ group is nearer to the healthy level, indicating the recovery of the link weights. 

To clearly observe the difference in link weights, we display the corresponding structural connectivity matrices in Figs.~\ref{figure1}(c-e), where the sparse connectivity matrix contains numerous short-range connections and fewer long-range connections. In the control individual, inter-hemispheric connections between homotopic regions of the same network are evident. In stroke patients, however, inter-hemispheric connectivity is significantly reduced. This is consistent with the findings by Griffis \textit{et al.}~\cite{siegel2016disruptions}, who demonstrated that the loss of inter-hemispheric connections, both structural and functional, is the predominant aberrant pattern observed in stroke.

Besides the distribution of weights, we also study the average values of different structural properties for the three groups and their corresponding standard errors of the mean (SEM).
In fact, the changes in the number of white-matter fibers are related to the fundamental network properties, such as average weight ($ \langle w \rangle $), average degree ($ \langle k \rangle $), maximum degree ($ \langle k_{max} \rangle $), degree assortativity ($ \langle r_c \rangle $), clustering coefficient ($ \langle C \rangle $), average path length ($ \langle L \rangle $), and network diameter ($ \langle D \rangle $).
These topological attributes of
structural networks play important roles in supporting potential communication processes in healthy and stroke brains~\cite{avena2018communication}. 
For example, Rocha \textit{et al.} found that the changes in the criticality regime appear to be strictly related to
changes of the stroke patient's network average degree and
connectivity disorder~\cite{rocha2022recovery}.  
Moreover, in contrast to the healthy brain, a typical stroke phenotype is characterized by a reduction in inter-hemispheric homotopic integration and a decrease in within-hemisphere segregation across different brain systems~\cite{siegel2016disruptions,siegel2018re}. Both of these alterations are indicative of a decline in network modularity, a phenomenon observed in stroke patients, even in the unaffected hemisphere~\cite{gratton2012focal}. To assess the degree of network segregation and integration in stroke, we employ modularity ($\langle Q \rangle$)and average path length ($\langle L \rangle$) as metrics in our work.

Figure~\ref{figure1}(f) shows the average weight $\langle w \rangle$, average degree $\langle k \rangle$, average path length $\langle L \rangle$,  and average modularity $\langle Q \rangle$ (community partition using the Louvain method~\cite{Blondel2008}), respectively. Table~\ref{table} reports the three groups' average values of considered structural properties. In addition, we perform independent $t$-tests to compare the significance of differences between the control group ($Con$) and two patient groups ($Pat. \ t_1$ and $Pat.\ t_2$) for each network property~\cite{lakens2013calculating}. At the same time, we calculate the absolute values of Cohen's $|d|$ to assess the effect size measure in statistics that quantify the difference between two group means in standard deviation units~\cite{lakens2013calculating}. A larger absolute value of Cohen's $|d|$ indicates a bigger effect size and the difference between groups.  
Table~\ref{table2} in the Appendix~\ref{Sup_T} shows very robust statistical tests.
All the results exhibit similar trends, where the architectures in the $ Pat. \ t_2 $ group are closer to the healthy control group than the ones in the $ Pat. \ t_1 $ group. 
In other words, the stroke will cause structural changes at three months post-stroke, and the network structure will recover gradually to some extent after one year. 

The difference observed in network properties between various groups may come from the changes in the number of white-matter fibers. Group analyses in stroke patients have revealed that the majority of lesions are affecting to a large part of white matter, which is critical in maintaining structural connectivity within cortical/subcortical regions~\cite{guggisberg2019brain,corbetta2015common}. Rocha \textit{et al.} also observed the changes in the number of fibers and topology in the stroke brain at a global level~\cite{rocha2022recovery}. At the level of the individual, there is more evidence of remodeling of white-matter connections (increasing fibers in existing connections) rather than recovery of white-matter connections~\cite{guggisberg2019brain,rocha2022recovery}.
Therefore, the recovery of structural connectivity from three to 12 months post-stroke may be closely related to the white-matter remodeling~\cite{guggisberg2019brain,rocha2022recovery}.

\renewcommand{\thetable}{\arabic{table}} 
\newcommand*{\thd}[1]{\multicolumn{1}{l}{#1}}
\begin{table}[b] 
	\centering
	\caption{Overview of the considered stroke brain networks. Columns are the index of group, the number of individuals in the group $N_g$, the average weight $\langle w\rangle$, average degree $\langle k\rangle$, average maximum degree $\langle k_{max} \rangle$, average clustering coefficient $\langle C\rangle$, average modularity $\langle Q\rangle$, degree assortativity $\langle r_c\rangle$, average path length $\langle L\rangle$, and average network diameter $\langle D\rangle$.}
	\begin{tabular}{*{10}{l}}
		\toprule
		\thd{Group}  & \thd{$N_g$} & \thd{$\langle w\rangle$} & \thd{$\langle k\rangle$} & \thd{$\langle k_{max}\rangle$} & \thd{$\langle C\rangle$} & \thd{$\langle Q\rangle$} & \thd{$\langle r_c\rangle$} & \thd{$\langle L\rangle$} & \thd{$\langle D\rangle$} \\
		\midrule
		$Con$     & 46     & 0.058                   & 17.94                   & 73.54                          & 0.45                   & 0.71                   & 0.17                     & 0.05                   & 1.30                  \\			
		$Pat. \ t_1$  &54       & 0.070                   & 15.27                   & 60.78                          & 0.46                   & 0.74                   & 0.19                     & 0.09                   & 1.71                   \\
		$Pat. \ t_2$   &59      & 0.065                   & 16.05                   & 66.98                          & 0.46                   & 0.73                   & 0.19                     & 0.08                   & 1.53                   \\
		\bottomrule
	\end{tabular}
	\label{table}
\end{table}

\subsection{Stroke-induced structural damages weaken the brain's activation propagation ability}\label{Adoption time}
The discoveries of structural alterations in the strokes call our great interest and motivate us to examine
their impacts on the brain's activation propagation ability. We aim to study how a weak perturbation spreads to the global network when it occurs. Therefore, following Ref.~\cite{mivsic2015cooperative}, we use the first adoption time
$t_{ij}$ to indicate brain function regarding activation propagation ability. It is defined as the time that node $j$ is first activated by a weak perturbation from a seed node $i$. We set $t_{ii}=0$ for the self-activated nodes, and $t_{ij} = t_{max}$ for a node $j$ can not be activated by a seed node $i$ within the maximum spread time $t_{max}$. 

Figures~\ref{figure2}(a-c) show the spread patterns for the three representative individuals in Fig.~\ref{figure1}(a) (i.e., No.~$12$ in $Con$, $ Pat. \ t_1 $ and $ Pat. \ t_2 $ group). Colors in each node represent the pace of the first adoption time under the same initial perturbation seed (indicated by the arrow). The blue node indicates that the adoption time is faster, while the red one is slower. The time from a specific seed node to others takes significantly longer for the individual in $ Pat. \ t_1 $, indicated by the more prominent red nodes. After one year, the adoption time for the stroke patient in $ Pat. \ t_2 $ shows a considerable improvement, comparable to a healthy individual's level.

To decrease the effect of the initial seed, in Figs.~\ref{figure2}(d-f), we investigate the adoption time matrices for the same individual in Figs.~\ref{figure2}(a-c). The matrix element represents the first activated time for a perturbation started at a particular seed node (rows) to reach another one (columns). We have arranged the matrices with ROIs based on the cortical parcellation of Gordon \textit{et al.}~\cite{gordon2016generation}. From the adoption time matrices of the representative individuals, we can observe that the patient at three months post-stroke has quite more red regions, implying that s/he needs more time to spread the activation. In other words, the stroke patient can not propagate the activation efficiently. 
Interestingly, the adoption time at $1$ year post-stroke brain gradually approaches the normal level in the healthy control group, implying the recovery of the ability to respond to stimuli and propagate activation to other brain regions. These results suggest that the brain undergoes gradual repair over time, corresponding with the statistical behaviors observed in brain structure as discussed in Section \ref{SC}. 

In the above discussion, we examine the brain's activation propagation ability at the individual level. However, do these phenomena persist across the entire group? To answer
this question, we analyze the adoption time matrices for each individual at different thresholds $\omega_c$. We first calculate the mean adoption time of individual $m$ over all elements in the matrix, $T_m^A$, excluding the elements with $t_{ij}=0$ and $t_{ij}=t_{max}$. Then, we can calculate the average spread time $\langle T^A \rangle$ over all the individuals in the same group with 
\begin{equation}
	\langle T^A \rangle = \frac{1}{N_g}\sum_{m=1}^{N_g}T_m^A,
	\label{eq:7}
\end{equation}
where $N_g$ is the number of individuals in the group (i.e., the sample size). We also obtain the corresponding standard error of the mean (SEM).
Figure~\ref{figure3}(a) shows the average adoption time across the whole brain as a function of the activation threshold $\omega_{c}$. At the same activation threshold, we can see that the healthy control group has the shortest average adoption time, meaning the speed of activation propagation is the fastest in the healthy brain. The slowest case happens in the three-month post-stroke group. The 1-year post-stroke group's adoption time was close to the healthy control group, indicating the recovery of the brain's ability for activation propagation.

\begin{figure}[!t]
	\centering
	\includegraphics[width=1\linewidth]{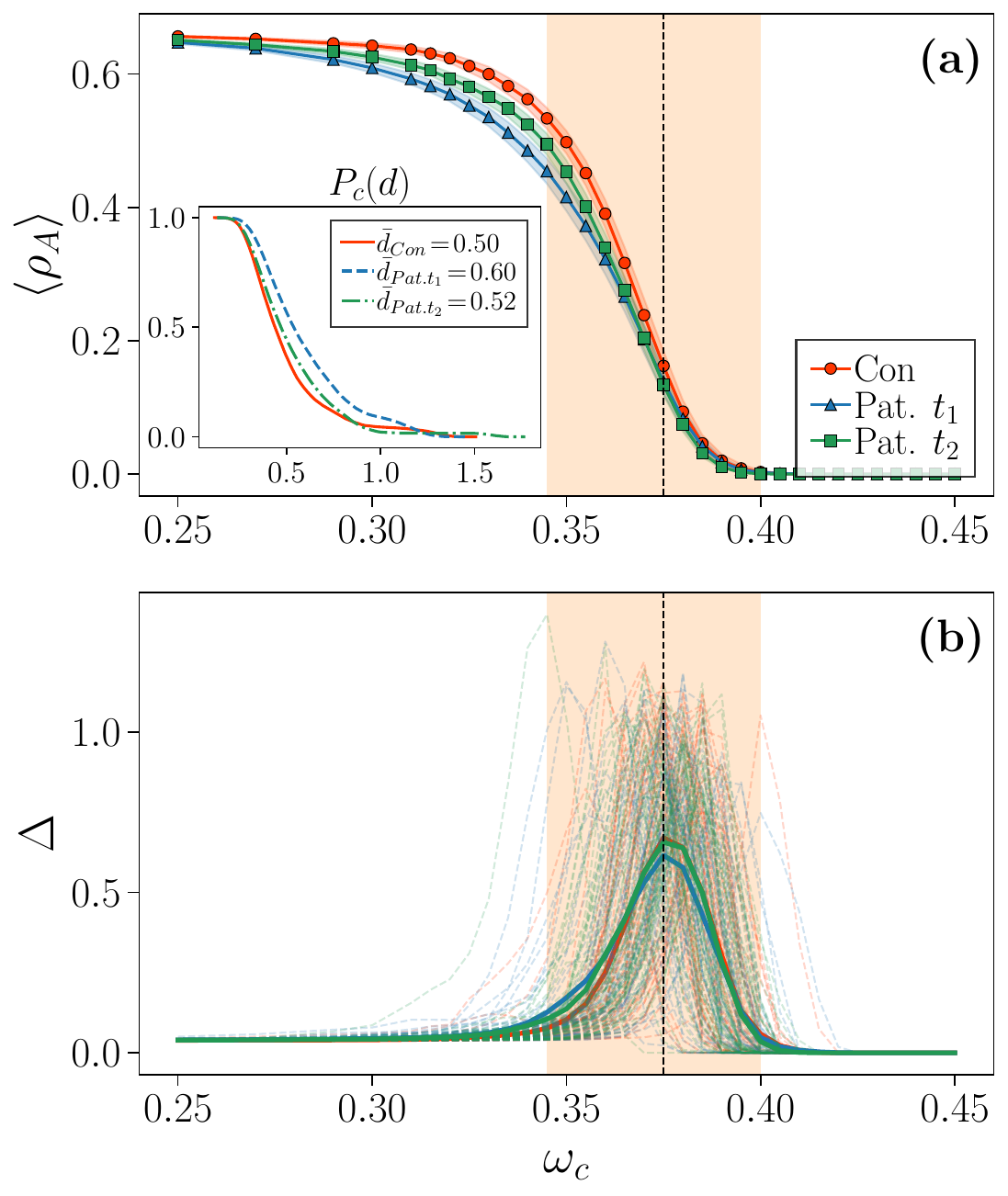}
	\caption{The average activation rate $\langle \rho_A \rangle$ and variability $\Delta$ as a function of the activation threshold for different groups. (a) $\langle \rho_A \rangle$ vs. $\omega_{c}$. The shaded regions on the symbols show the corresponding SEM around the expected value. The inset in (a) shows the complementary cumulative distribution functions of the Euclidean distances $d$, $P_c(d)$, for variable $\langle \rho_A \rangle$. The mean values of each $d$ are reported in the legend. 		
		(b) $\Delta$ vs. $\omega_{c}$. The solid lines represent the group's average and each dashed line indicates the individual case. The red, blue, and green dashed line  represents the individual in $Con$, $Pat. \ t_1$ and $Pat. \ t_2$, respectively. The light orange shadow area marks the range of the critical points of $\omega_{c}$ for all control and stroke individuals. The vertical dashed lines indicate the case of the average critical activation threshold $\omega_{c} = 0.37$.} 
	\label{figure4}
\end{figure}
\begin{figure*}[!ht]
	\centering
	\includegraphics[width=1.0\textwidth]{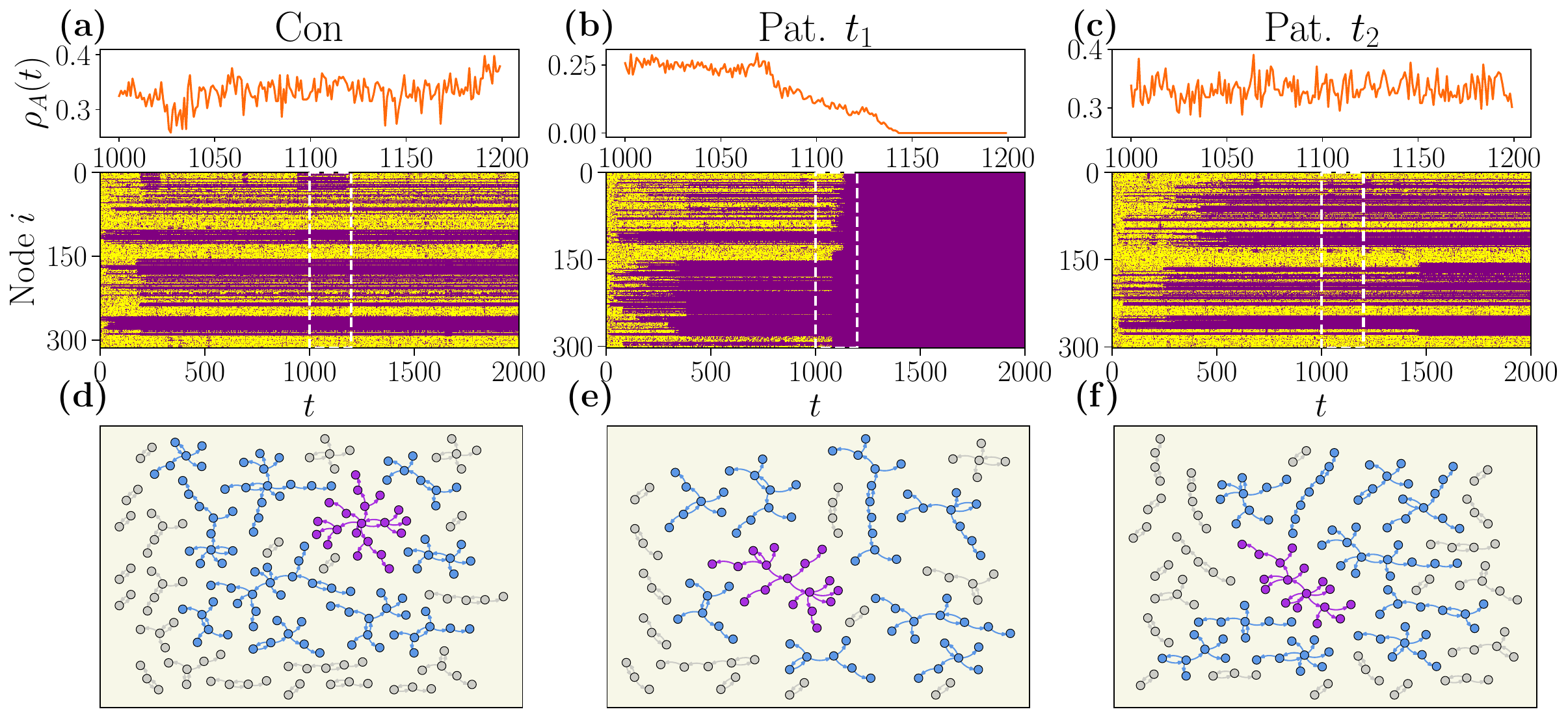}
	\caption{The self-sustaining oscillatory patterns (a-c, Bottom) and corresponding core networks (d-f) for the three representative individuals in Figs.~\ref{figure2}(a-c).
	In the top panels of (a-c), we show the trajectories of the activation fraction $\rho_A(t)$ in time window $t=[1000,1200]$ as marked with white dashed boxes at the bottom panels. Nodes with purple, blue, and gray in (d-f) indicate the core subnetworks' maximum, moderate, and minor sizes (i.e., the number of nodes in the subnetworks), respectively. Each directed link depicts the main pathway of activation propagation.}
	\label{figure5}
\end{figure*}
\begin{figure*}[!ht]
	\centering
	\includegraphics[width=1\textwidth]{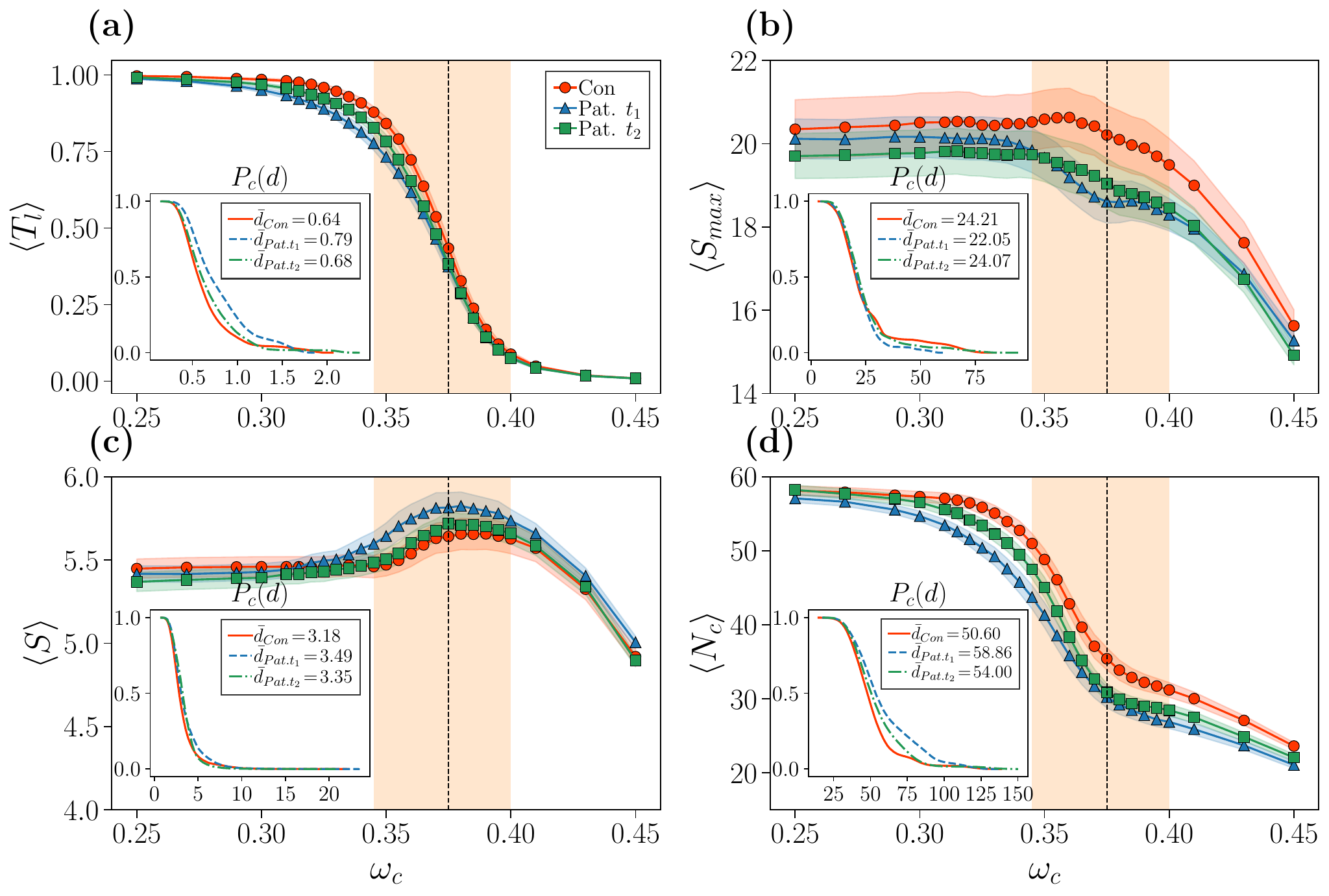}
	\caption{The average lifetime and properties of core networks on the self-sustaining oscillation patterns for different groups. (a) The average lifetime of self-sustaining oscillation patterns $\langle T_l\rangle$, (b) the average largest core network size $\langle S_{max} \rangle$, (c) the mean core network size $\langle S\rangle$, (d) the average number of core subnetworks $\langle N_c\rangle$,  as a function of the activation threshold $\omega_{c}$. The insets show the complementary cumulative distribution functions of the Euclidean distances $d$, $P_c(d)$, for each variable. The mean values of each $d$ are reported in the legend. The shaded regions on the symbols show the corresponding SEM around the expected value. The light orange shadow area marks the range of the critical points of $\omega_{c}$ for all control and stroke individuals. The vertical dashed lines indicate the case $\omega_{c} = 0.37$.}
	\label{figure6}
\end{figure*}

This phenomenon observed at the whole-brain level prompts the question: does it also occur at different scales within the brain? To investigate this, in Figs.~\ref{figure3}(b-d) we separately analyze the average adoption time within the left or right hemispheres, between the left and right hemispheres, and within specific brain regions. At the hemispheric scale, as shown in Figs.~\ref{figure3}(b) and (c), the results for the average adoption time in intra-hemispheres and inter-hemispheres are consistent with those observed at the whole-brain level. 
Furthermore, at a smaller scale, we examine the average adoption time of different brain regions for each group with fixed $\omega_{c}=0.03$. In the bottom of Fig.~\ref{figure3}(d), we plot each group's average adoption time and corresponding SEM with seven considered brain regions. We also highlight the nodes in the same cortical parcellation with red color at the top of Fig.~\ref{figure3}(d). 
Table~\ref{table3} in the Appendix~\ref{Sup_T} reports robust statistical tests about the significance and effect sizes.
The average adoption time follows the same behavior seen at the whole-brain and hemispheric scales. This phenomenon indicates that the ability to propagate activation in stroke patients is weakened across different brain scales. After one year, there is a recovery to some extent, but the activation propagation ability still lags behind that of control individuals.

\subsection{The stroke weakens the brain's self-sustaining oscillation ability}\label{TLSOP}

Besides the activation propagation ability, we are also interested in the dynamic patterns of activation propagation. Here, we adopt the same model to study the brain's rhythm-related self-sustaining oscillation patterns. In the healthy brain, Huo \textit{et al.}~\cite{huo2022time} have observed the generation of time-limited self-sustaining oscillatory patterns with varying lifespans, where the patterns depend on the seeds initially selected and the activation threshold. As the brain lesion locations caused by stroke vary across individuals~\cite{corbetta2015common}, different initial perturbation sources (i.e., initial activation seeds) would result in various patterns of self-sustaining oscillations. 
Therefore, we activate all nodes at the initial moment to reduce the initial activation seeds' impact and better compare the patterns between healthy and stroked individuals.

To better observe the patterns of self-sustaining oscillations, we need to identify a reasonable activation threshold $\omega_{c}$. We firstly denote an instantaneous activity $\rho_A(t)$ as the fraction of activation nodes at time step $t$, which can be calculated as  
\begin{equation}
\rho_A(t)=\frac{1}{N}\sum_{i=1}^N S_i(t),	 
	\label{eq:rhoAt}
\end{equation}
where $N$ is the number of nodes.
Then, we can obtain the mean stationary proportion of activation nodes in the whole brain (i.e., the average activation rate or activity range) as 
\begin{equation}
	\langle \rho_A \rangle =\frac{1}{t_s-t_r}\sum_{t=t_r}^{t_s} \rho_A(t),  
	\label{eq:ave_rhoA}
\end{equation}
where $t_s=2000$ is the simulated total time and $t_r=1000$ is the initial transient time (i.e., we
discard the initial transient dynamics with the first 1000 time steps). Figure~\ref{figure4}(a) illustrates the association between the average activation rate $\langle \rho_A \rangle$ and the activation thresholds $\omega_{c}$ for different groups. All the results were obtained by $1000$ realizations. This figure shows that at most activation thresholds, the activity range in the network of stroke group is smaller than in the control group. 
This result indicates that structural alterations are associated with smaller activation propagation range within the model.

We next use the Euclidean distance, $d$, to quantify the variability between a given simulated variable and the corresponding control group average~\cite{rocha2022recovery} (see Eq.~(\ref{eq:variability}) in Appendix \ref{variability}).  
This measure considers the dynamic variables' behavior across all activation threshold values $\omega_{c}$. A low $d$ indicates minimal variability in dynamic parameters relative to controls. In contrast, a high $d$ reflects significant variability of the parameters across participants and time points for strokes, suggesting the presence of abnormal dynamics in the group. The inset in Fig.~\ref{figure4}(a) shows the complementary cumulative distribution functions of the Euclidean distances $d$, $P_c(d)$, for $\langle \rho_A \rangle$. This figure shows higher variability in stroke than controls, and higher variability at $Pat.\ t_1$ than $Pat.\ t_2$ as reported the mean values of each $d$ in the legend.


With the average activation rate $\rho_A$ on hand, we can determine a reasonable range of activation threshold $\omega_{c}$ based on the critical brain hypothesis. It is hypothesized that the healthy brain works close to a phase transition (critical point), which supplies optimal conditions for activation spreading and response to inputs~\cite{haimovici2013brain,rocha2022recovery}. Inspired by the critical brain hypothesis, we use the variability $\Delta$ to numerically determine the critical point of $\omega_{c}$~\cite{crepey2006epidemic,yang2023asymmetric}, where $\Delta$ is measured as 
\begin{equation}
	\Delta = \frac{\sqrt{{\langle \rho_A^2\rangle}-{\langle \rho_A\rangle}^2}}{{\langle \rho_A\rangle}}, 
	\label{eq:8}
\end{equation}
where $\langle \rho_A^2\rangle=\frac{1}{t_s-t_r} \sum_{t=t_r}^{t_s}\rho_A^2(t)$.
Notice that there is peak at a critical point for the variability $\Delta$~\cite{crepey2006epidemic,yang2023asymmetric}. Therefore, we numerically estimated the critical $\omega_{c}$ by determining the place of the variability's peak. Figure~\ref{figure4}(b) shows the relationship between $\Delta$ and the activation threshold for each group, where the solid line represents the group's average and each dashed line indicates the individual case. The red, blue, and green dashed line represents the case of each individual in $Con$, $Pat. \ t_1$ and $Pat. \ t_2$, respectively. The peaks of these curves indicate the critical point for each individual. The shadow area marks the range of the critical points for all control and stroke individuals. The average critical activation thresholds for the three groups are roughly the same, around $\omega_{c} \sim 0.37$. 

A possible cause of no significant group differences on the critical activation threshold is that it is related to the severity of stroke-induced damage. We can see that not all individuals exhibit changes in $\omega_c$
following a lesion; while some patients' critical activation thresholds remain comparable to those of controls, others show significant deviations from normality. To verify whether the critical activation threshold $\omega_c$
is associated with the severity of stroke-induced damage, we extend an artificial stroke model with adjustable levels of damage applied to the healthy connectomes~\cite{janarek2023investigating} (see the artificial stroke model in Appendix~\ref{ART} for more details). The results in Fig.~\ref{comment3} in the Appendix~\ref{ART} report a good qualitative agreement between the outputs of artificial and real stroke networks. Specifically, when the severity level is at $15\%$ and $10\%$ (corresponds to the level in $Pat. \ t_1$ and $Pat. \ t_2$), there are no significant group differences on the critical activation threshold $\omega_c$. However, we observe a leftward shift of critical $\omega_c$ in a more serious damage level at $30\%$. These findings suggest that the lack of significant group differences on the critical activation threshold might result from less severe stroke-induced damage.

Next, we will use the group's critical activation threshold $\omega_{c} \sim 0.37$ to investigate the self-sustaining oscillations. In the bottom panels of 
Figs.~\ref{figure5}(a-c), we show the self-sustaining oscillatory patterns for the three representative individuals in Figs.~\ref{figure2}(a-c). We also show the time evolution of the activation fraction $\rho_A(t)$ in time window $t=[1000,1200]$ in the top panels of Figs.~\ref{figure5}(a-c).
It is easy to observe the emergence of collective oscillations from the trajectories of $\rho_A(t)$ in the whole brain.
Additionally, from the self-sustaining patterns shown in the bottom of Figs.~\ref{figure5}(a-c), we observe that the activities of different nodes are sequential or alternating throughout the evolutionary process. Specifically, each node's time series of activity is not continuous but exhibits intermittent periods of inactivity. This behavior suggests that the activities propagate among the nodes, leading to a limit-cycle-like dynamics, characteristic of oscillatory behavior~\cite{huo2022time}. 

Moreover, it is easy to observe that Figs.~\ref{figure5}(a) and (c) exhibit a phenomenon of sustained activity propagation (i.e., typical self-sustained patterns). Interestingly, the self-sustaining oscillatory pattern in Fig.~\ref{figure5}(b) is time-limited, where it experiences a sudden cessation of activity for all nodes during the propagation process. It should be noted that the time-limited self-sustaining oscillatory pattern is not unique for the individuals in $Pat.\ t_1$. This kind of time-limited pattern also appears in $Con$ and $Pat.\ t_2$ due to individual differences in critical thresholds as shown in Fig.~\ref{figure4}(b) (see the patterns for all individuals in Figs.~\ref{figureS2_con}-\ref{figureS2_pat_t2} in Appendix
\ref{A}). Strikingly, comparing with Fig.~\ref{figure5}(b) and (c), we find that the same stroke individual can recover from an un-sustained state to sustain, which implies the survival time of each node is longer for the individuals in $Pat.\ t_2$ than in $Pat.\ t_1$. 

Inspired by the self-sustained patterns in Figs.~\ref{figure5}(a-c), we next focus on the average lifetime
on the group level. Let's denote $t_{i}^{initial}$ and $t_{i}^{last}$ as the time points of a node $i$ initial and last activity in the simulated total time $t_s$, respectively. Then,  $t_i=t_{i}^{last}-t_{i}^{initial}$ is the time for a node $i$ that has passed between the first and last activity in the spreading process during the simulated total time $t_s$. In our case, as we active all the nodes initially, so $t_{i}^{initial}=0$ for all nodes, and we only need to find out the time point of node $i$ last activity, $t_{i}^{last}$, during the simulated total time $t_s$ ($t_s=2000$ as before).  
Finally, we define the lifetime of an individual $l$ activation propagation as the normalized survival time over all nodes. It can be written down 
\begin{equation}
	T_l = \frac{1}{N}\sum_{i=1}^{N}\frac{t_i}{t_s}, 
	\label{eq:9}
\end{equation}
where $ N $ is the number of nodes. $T_l\to1$ means the lifetime of an individual activation propagation is long, while $T_l\to0$ indicates short. Therefore, the average lifetime $\langle T_l\rangle$ over all individuals in the same group reflects the brain's self-sustaining oscillation ability at the group level. 
Figure~\ref{figure6}(a) shows the average lifetime $\langle T_l\rangle$ for the three groups at different activation thresholds. We observe that within the critical region where the brain operates (marked in the light orange shadow region), the average lifetime of the healthy control group is relatively longer than that of the stroke groups. For example, when $\omega_c=0.37$, $\langle T_l\rangle$ is 0.5381, 0.4634, and 0.4798 in $Con$, $Pat. \ t_1$ and $Pat. \ t_2$, respectively. This phenomenon indicates that stroke indeed impairs the brain's self-sustaining oscillatory ability. Moreover, the brain at one year post-stroke exhibits some recovery compared to the group three months post-stroke, which is consistent with the observation in Figs.~\ref{figure5}(b) and (c). Therefore, the brain's self-sustaining oscillation ability is weakened in stroke patients, but this ability will recuperate to some extent with the brain's healing.

\textbf{The dominant activation
paths and core network.} As we know, the lifetime of self-sustaining patterns is exponentially related to the size of the core network~\cite{huo2022time}, where the directed edges indicate the dominant activation paths that signal travel in the brain. It has been shown that the multiscaled rhythms are associated with the core network's size in the brain network~\cite{huo2022time}. An interesting question arises: how does the core network change in the stroke patient? To figure out the answer, we will re-introduce the approach of dominant activation
paths~\cite{huo2022time} and analyze the core network comprised
by the dominant activation paths in stroke brains (see Appendix ~\ref{dominant} for more details). Let's define an activation matrix $ M $, where each element $ M_{uv} $ represents the frequency of the activated connection $ v \rightarrow u $. Initially, we set $ M_{uv} = 0 $. For the evolution process involving node $u $ and one of its neighbor $ v $, if $S_u(t-1) = 0 $, $ S_u(t) = 1 $ and $ S_v(t-1) = 1 $, then we consider the connection $ v \rightarrow u$ as an activated connection at time step $t$ and update $ M_{uv} = M_{uv} + w_{uv} $. With this method, we obtain the cumulative $ M_{uv} $ over a time period. Next, we select the greatest $ M_{uv}$ from all $u$' neighbors in the self-sustaining pattern, i.e., the dominant one. Ultimately, we add a directed connection from the dominant node $v$ to node $u$. By doing so, every node $u$ has only one incoming connection from the dominant node $v$.
Following this identification method, we obtain a core network of dominant activation paths in which directed links represent the main pathways of activation propagation. 

In Figs.~\ref{figure5}(d-f), we show the core networks for the three representative individuals in Figs.~\ref{figure2}(a-c). 
It is easy to see that the size of the core network for the patient at three months post-stroke is smaller compared to the other two individuals, and the number of core subnetworks is also fewer. This preliminary observation suggests that brain damage caused by stroke decreases the dominant activation paths and weakens the brain's self-sustaining oscillation ability. To reduce the effects of individual differences, we quantify each individual's core network and analyze the core networks' properties at the group level. 
We show our results of the average maximum core network size $\langle S_{max} \rangle$, the mean core network size $\langle S\rangle$, and the average number of core subnetworks $\langle N_c\rangle$ as a function of the activation threshold $\omega_{c}$ in Figs.~\ref{figure6}(b), (c) and (d), respectively. We find that within the critical region(marked by the light orange shadow), the properties of core networks in the three-month post-stroke group immensely depart from the healthy group, while the case for the one-year post-stroke group is close to the healthy one. These phenomena imply that the lifetime and recovery of self-sustaining patterns are related to the properties of core networks.

\section{discussion and conclusion}\label{IV}
Healthy brains usually display highly efficient activation propagation and self-sustained oscillation abilities, but how the brain network alters these dynamic behaviors after a stroke remains unclear. In addition, the recovery of structure and dynamic behaviors after stroke over time still needs to be clarified precisely. 
In this study, we analyzed the real structural connectivity networks of stroke and 
revealed that stroke changes the network properties, such as connection weights, average degree, average clustering coefficient, community, etc. In addition, we adopted a reaction-diffusion model to investigate stroke patients' activation propagation and self-sustained oscillation abilities. We found that the stroke decreased the brain's activation propagation efficiency and self-sustaining oscillation ability. More specifically, the stroke patients' brains decreased the speed of activation spreading, lowered the activation rate, reduced the lifetime of self-sustained oscillatory patterns, and weakened the sizes and numbers of core networks. Most importantly, we observed the loss of activation spreading and self-sustained oscillation abilities at three months post-stroke while recovering after one year, driven by the structural connection repair. Our work may enable us to understand better the association between architecture and function in brain disorders.

We performed our model on the structural connectivity matrix 
with node-wise normalization of connectivity weights~\cite{rocha2022recovery,rocha2018homeostatic}, a mechanism of homeostatic plasticity~\cite{deco2014local,hellyer2016local,abeysuriya2018biophysical}. As discussed in Ref.~\cite{rocha2022recovery,rocha2018homeostatic}, this simple adjustment optimized the macroscopic dynamics, thereby enhancing the robustness of critical transitions. A key characteristic of node-wise normalization is that it minimizes the variability of the critical points and neuronal activity patterns among healthy participants. 
Therefore, by node-wise normalization, we can use the identical activation threshold $\omega_c$ to compare the dynamic behaviors between healthy participants and strokes, and stroke patients at different time points. 
However, it is not clear whether other normalization methods on the connectivity weights will affect our findings. As the original data from Ref.~\cite{rocha2022recovery} has been performed node-wise normalization on the connectivity weights, we can not investigate the influences of different normalization methods on the spreading dynamic behaviors between controls and strokes. But, we can test the sensitivity of the choice of normalization methods with the data from our previous work~\cite{zheng2020geometric} (see the data and results in Appendix~\ref{B}). The simulated results on the new dataset imply that our findings could be robust for choosing normalization methods if we rescaled the activation threshold to the same level. However, testing the sensitivity of the normalization methods on the spreading dynamics between patients and strokes is not our main goal in this work. We will investigate it with more empirical data in future work.

Models of large-scale neuronal dynamics are essential for understanding and predicting neuronal activity at the macroscopic level~\cite{breakspear2017dynamic}. However, the mechanisms governing large-scale communication processes across the whole brain remain poorly understood~\cite{breakspear2017dynamic}. We adopted a biophysically-inspired whole-brain model to explore the dynamic behavior of activation propagation in healthy controls and strokes. Although our model is simple, it can capture the differences brought about by structural connections between healthy participants and strokes, as well as stroke patients at different time points. A limitation of this work is that our model can not directly reveal the mechanisms underlying stroke recovery, including the potential role of increased activity in specific brain regions. This issue has yet to be thoroughly investigated or discussed. 

In addition, it is not clear why some seeds can not trigger activity across the network. As we know, the activity propagation behavior depends on the seeds and activation threshold. For a given activation threshold, the attributes of a network node play key roles in trigger activity propagation. We examined the relationship between the degree of the seed node and the average activation rate $\langle \rho_A \rangle$ (i.e., spread range) and found that the seeds with small degrees will lower the average activation rate (not shown here).
But they are not linearly dependent. This result implies that besides the degree of nodes, there are some other essential attributes affecting the activity propagation in the human brain networks. We are highly interested in this issue and will examine it in the future.



\appendix

\section{The influence of normalized connectivity weights}
\label{B}
Here, we want to test whether other normalization methods on the connectivity weights will affect our results. It is a pity that the original data from Ref.~\cite{rocha2022recovery} has been performed node-wise normalization on the connectivity weights. Thus, we can not directly investigate the influence of different normalization methods on the dynamics between patients and strokes. However, we can test the sensitivity of the choice of normalization methods with the data in healthy participants from our previous work~\cite{zheng2020geometric}. The new dataset has $40$ subjects and $N=1014$ nodes for each structural connectivity matrix. The weights of the edges represent the fiber densities and have not been normalized by any method. 

\begin{figure*} 
	\centering
	\includegraphics[width=0.8\textwidth]{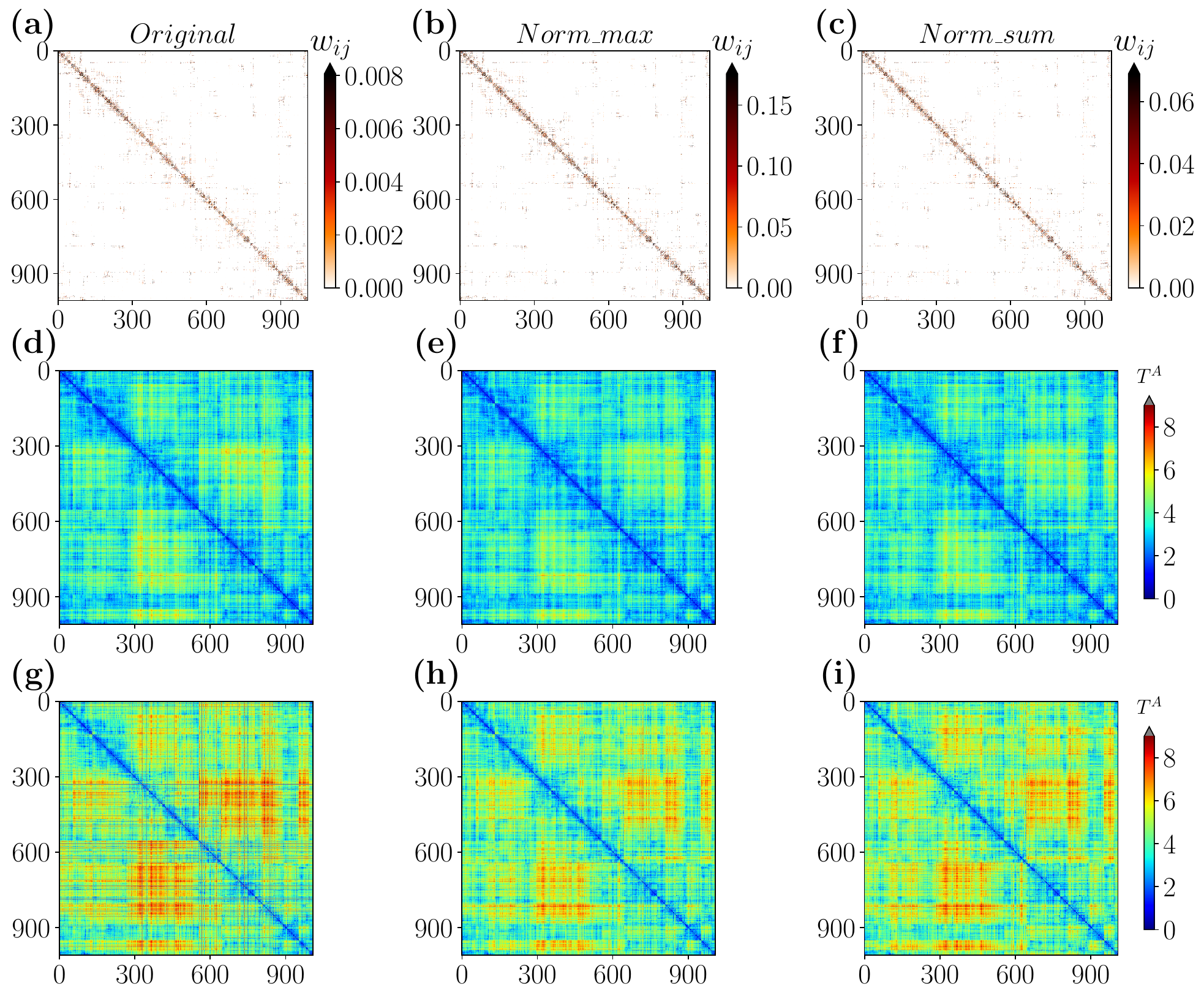}
	\caption{(a-c) The structural connectivity matrices for a representative individual with original edge weights, normalized by maximum weight and by the sum of the edge weights of neighbors, respectively.
		(d-f) The adoption time matrices calculating on the corresponding structural connectivity matrices (a-c), where the $\omega_c/{\langle s \rangle}=0.01$. (g-i)   
		same as (d-f) but fixed $\omega_c/{\langle w \rangle}=1.00$. The adoption time matrices for different normalization methods on the connectivity weights are qualitatively similar when we rescaled the activation threshold $\omega_c$ by average strength $\langle s\rangle$ or average weight $\langle w\rangle$}.
	\label{comment1}
\end{figure*}

\begin{figure*} 
	\centering
	\includegraphics[width=0.65\textwidth]{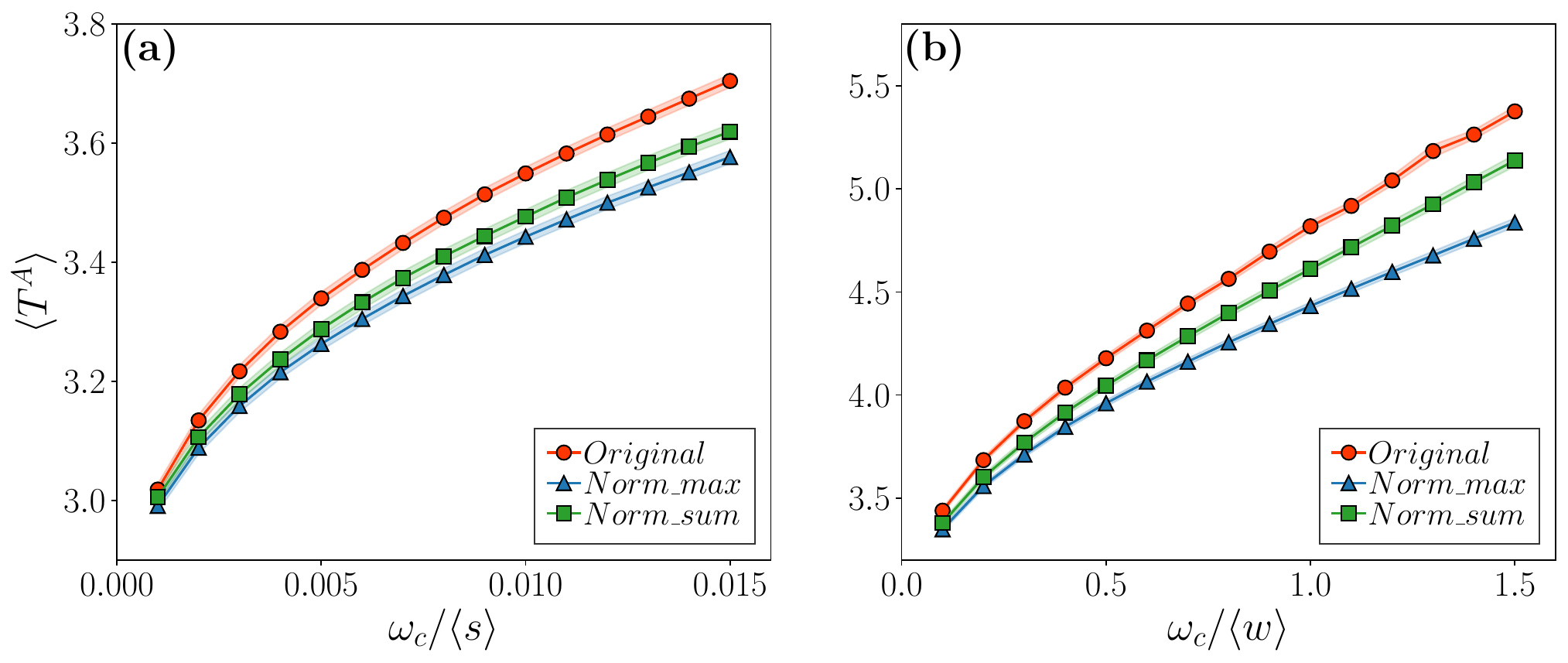}
	\caption{The average adoption time on the new dataset with different normalization methods on the connectivity weights. (a) and (b) show the average adoption time as a function of rescaled activation threshold $\omega_c/\langle s\rangle$ and $\omega_c/\langle w\rangle$, respectively. The shaded regions on the symbols show the corresponding SEM around the expected value. The average adoption time $\langle T^A\rangle$ vs the rescaled activation threshold $\omega_c/\langle s\rangle$ and $\omega_c/\langle w\rangle$ follow the same trends, implying that our results may be robust for choosing normalization methods if we rescaled the activation threshold to the same level.}
	\label{comment2}
\end{figure*}

We mainly focus on two normalization methods.
One is normalizing by maximum weight (naming $Norm_{-}{max}$), and the other is normalizing 
by the sum of the edge weights of neighbors (naming $Norm_{-}sum$), i.e., node-wise normalization with Eq.(\ref{eq:1}). In Figs.~\ref{comment1} (a)-(c), we show the 
structural connectivity matrices for a representative individual with original edge weights, normalized by maximum weight and by the sum of the edge weights of neighbors, respectively. 
As shown in Figs.~\ref{comment1} (d)-(i), the results of the adoption time matrices for different normalization methods on the connectivity weights are qualitatively similar when we rescaled the activation threshold $\omega_c$ by average strength $\langle s\rangle$ or average weight $\langle w\rangle$. In addition, as shown in Fig.~\ref{comment2}, the average adoption time $\langle T^A\rangle$ vs the rescaled activation threshold $\omega_c/\langle s\rangle$ and $\omega_c/\langle w\rangle$ in 
the connectivity matrix with original, normalized by maximum weight, and normalized by the sum of the edge weights of neighbors follow the same trends. These results imply that our results may be robust for choosing normalization methods if we rescaled the activation threshold to the same level.


\section{Quantify individual variability relative to controls}
\label{variability}
We quantify the similarity between a given simulated variable and the corresponding control average using the Euclidean distance $d$~\cite{rocha2022recovery}:
\begin{equation}
	d=\sqrt{\sum_{\omega_{c}}(\langle X\rangle-X_g(\omega_{c}))^2},
	\label{eq:variability}
\end{equation}
where $X_g(\omega_{c})$ is a given variable at different group $g$ and activation threshold $\omega_{c}$, while $\langle X\rangle$ is the corresponding controls average. 
This measure incorporates the behavior of dynamic variables across all values of the activation threshold 
$\omega_{c}$. A low $d$ indicates minimal variability in dynamic parameters compared to controls, whereas a high 
$d$ signifies substantial variability in the variables across participants and patients from different time points, which suggests the presence of abnormal dynamics within the group.

\section{Artificial stroke model}
\label{ART}

To better understand why there is no significant effect of group differences on the critical activation threshold, $\omega_c$. We extend an artificial stroke model with adjustable levels of damage applied to the healthy connectomes~\cite{janarek2023investigating}. The basic idea is to exame whether the critical activation threshold 
$\omega_c$
is associated with the severity of stroke-induced damage.
Starting from a healthy empirical connectome, the artificial stroke model disrupts interconnections between distinct brain regions. Specifically, we randomly select a fixed fraction of nodes—representing the severity of the stroke—and sever their connections to neighboring nodes outside their own region. This approach preserves the internal structure of nodes within the same region while effectively reducing the connectivity between different brain regions. 

We set the fraction of nodes as $15\%$ and $10\%$ disconnected from the rest of the brain regions, which results in the artificial stroke networks having almost the same mean degree in $Pat.\ t_1$ and $Pat.\ t_2$, respectively. For comparison purposes, we also choose the fraction of nodes as $30\%$ to represent a more serious case. After disrupting the interconnections, we re-normalize the weights in connectivity matrices with Eq.~(\ref{eq:1}), thereby shifting the critical activation threshold to the same level. For each healthy connectome and a given stroke severity, we generate $10$ artificial stroke networks and then calculate the average activation rate $\rho_A$ and variability $\Delta$. 

\begin{figure}[!t]
	\centering
	\includegraphics[width=0.48\textwidth]{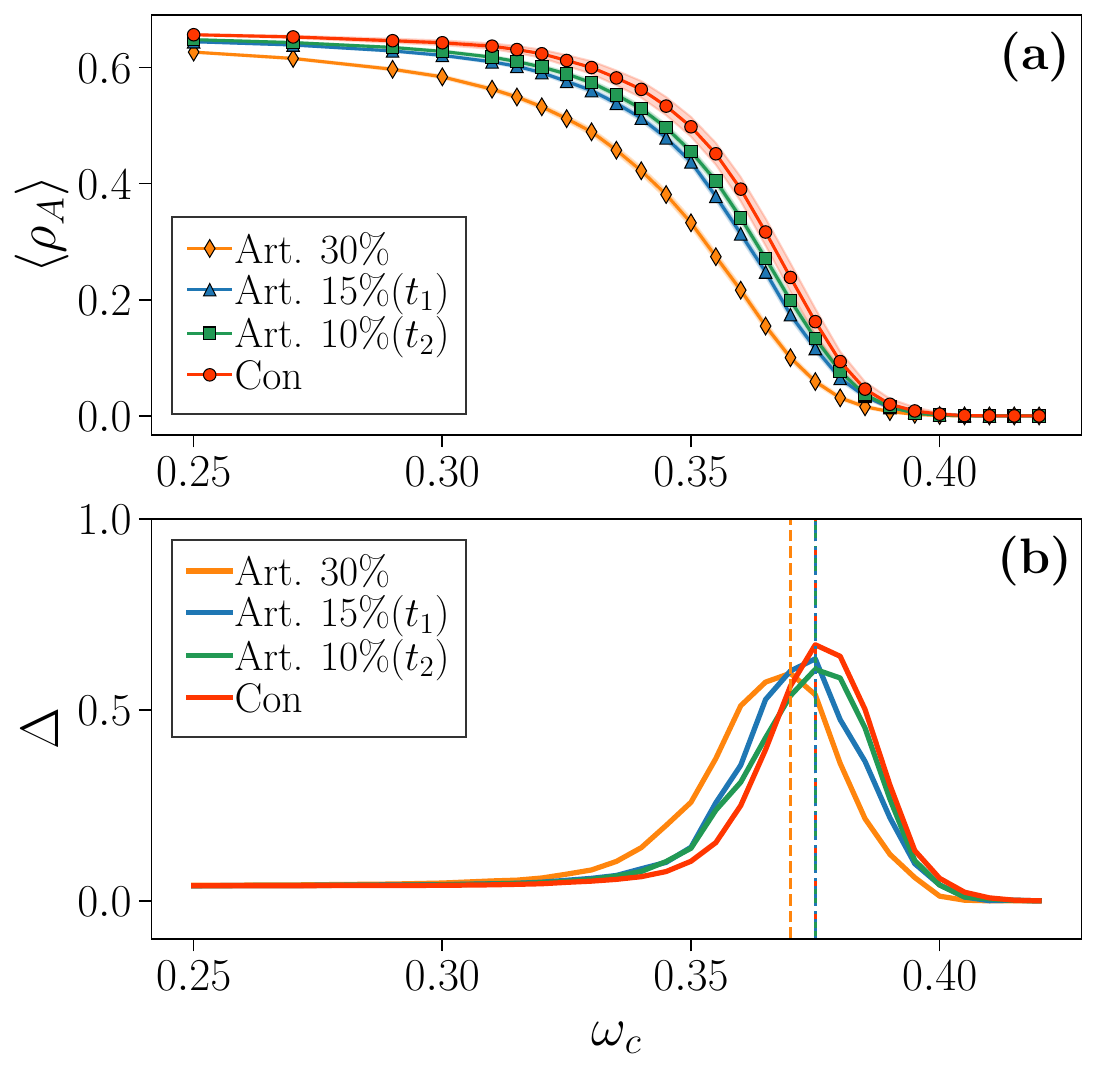}
	\caption{The average activation rate $\langle \rho_A \rangle$ and variability $\Delta$ as a function of the activation threshold $\omega_c$ for varying the severity in artificial stroke networks. The vertical dashed lines indicate the critical points of $\omega_{c}$ for different stroke severity.}
	\label{comment3}
\end{figure}

Figure~\ref{comment3} illustrates similar results between the artificial and real stroke networks. Specifically, at severity levels of $15\%$ and $10\%$, no significant group differences in the critical activation threshold 
$\omega_c$ 
are observed. However, at a higher damage level of $30\%$, a leftward shift in the critical 
$\omega_c$
is evident. These findings suggest that the absence of significant group differences in the critical activation threshold may be attributed to the relatively mild severity of stroke-induced damage.


\section{Approach of dominant activation paths}
\label{dominant}

The central concept behind identifying the dominant activation pathways is derived from the method of dominant phase-advanced driving (DPAD)~\cite{qian2010structure,liao2011oscillation}, which has been pivotal in demonstrating that the presence of a DPAD loop is a key feature of self-sustained patterns ~\cite{mi2013long,xu2014simplified,xu2013controlling,roxin2004self}. The core idea of DPAD is to identify a loop of connected nodes where each non-oscillatory node can begin to oscillate only if it is driven by one or a few oscillatory interactions with advanced phases. Once such a DPAD loop is identified, the oscillations within this loop propagate throughout the system, giving rise to a periodic, self-sustained oscillatory pattern. However, since self-sustained patterns in our context are irregular rather than periodic, the DPAD loop approach is not directly applicable.

Inspired by the principles of DPAD, we re-introduce a method called the dominant activation paths~\cite{huo2022time}. The key idea is to construct an activation matrix $M$, where each element $M_{uv}$ represents the frequency of the activated connection from node $v$ to node $u$. As an inactivated node $u$ node becomes activated if and only if its activated neighbors drive it, the contribution of each activated neighbor $v$ is 
reflected in $M_{uv}$. In other words, $ M_{uv} = M_{uv} + w_{uv}$ is updated if $S_u(t-1) = 0 $, $ S_u(t) = 1 $ and $ S_v(t-1) = 1 $. Note that the activation matrix $M$ is derived from statistical frequencies of activations, rather than from the interaction that provides the most significant contribution at a time step. In this way, we can get the accumulated $M_{uv}$ for a time period. For each node $u$, we identify the maximum $M_{uv}$ across all its neighboring nodes, i.e., the dominant contribution. Then,
we put a directional link from the dominant node $v$ to the
node $u$. Ultimately, every node $u$ has a single incoming connection from its dominant node $v$. All these
incoming links will form the core subnetworks. 
Figures~\ref{figure5}(d)-(f) are the network visualizations of activation matrix $M$ that only consider the maximum element $M_{uv}$.

\section{Supplementary Figures}
\label{A}

\begin{figure}[!t]
	\centering
	\includegraphics[width=0.45\textwidth]{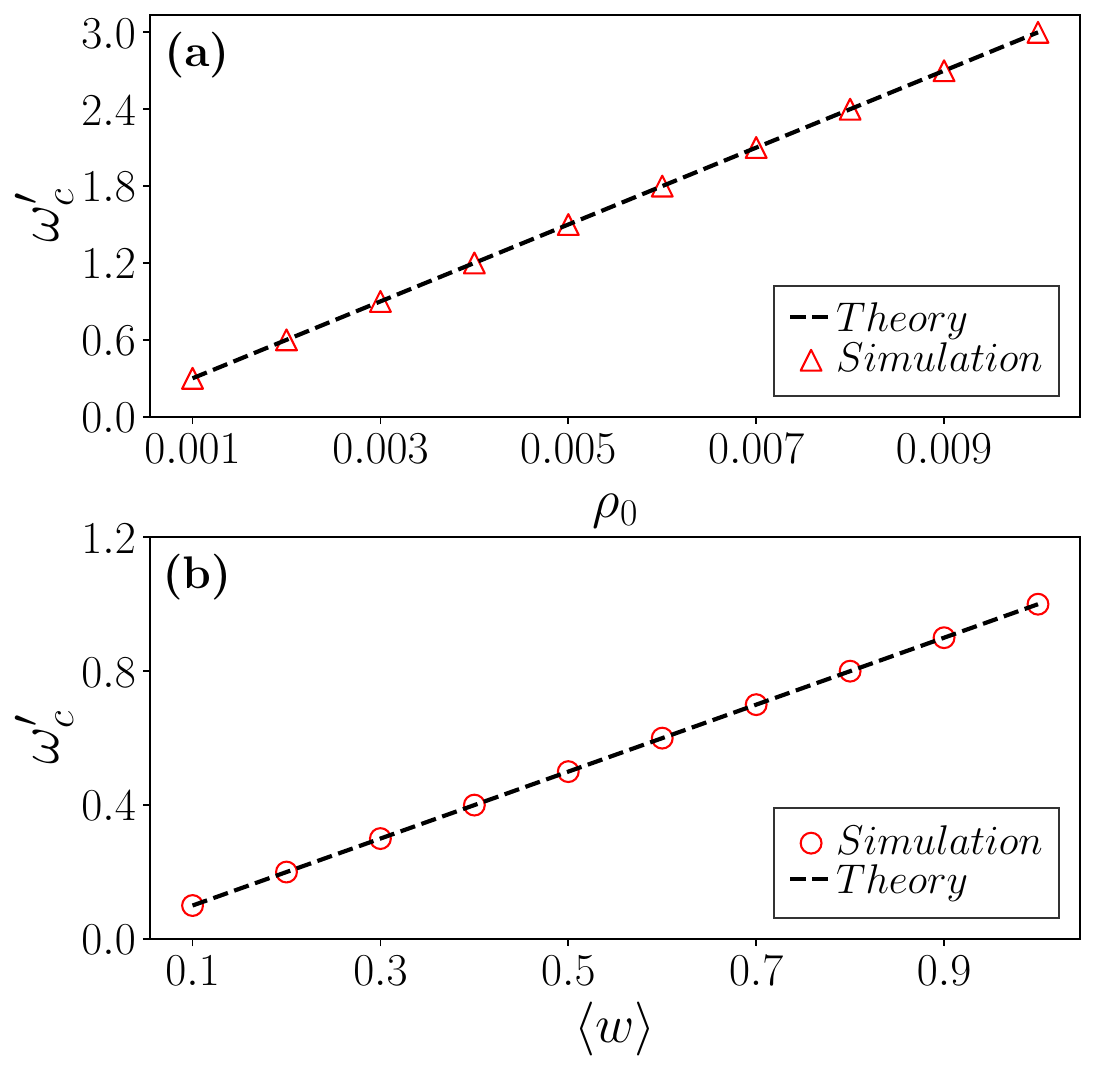}
	\caption{Analytic approximation vs simulations. (a) the critical point $\omega'_c$ vs the initial activation ratio $\rho_0$ for fixed $\langle w \rangle=0.3$. (b) the critical point $\omega'_c$ vs average weight $\langle w \rangle$ for fixed $\rho_0=0.001$. We simulate the activation propagation model on synthetic fully connected networks with uniform connection weights. The opened symbols are the simulation results, and the dashed lines stand the theoretical analysis from Eq.~($\ref{eq:condition2}$). The network size is fixed as $N=1000$. All the results have been averaged over $10^3$ independent realizations. }
	\label{figureS1}
\end{figure}

\begin{figure*}[!h]
	\centering
	\includegraphics[width=1\textwidth]{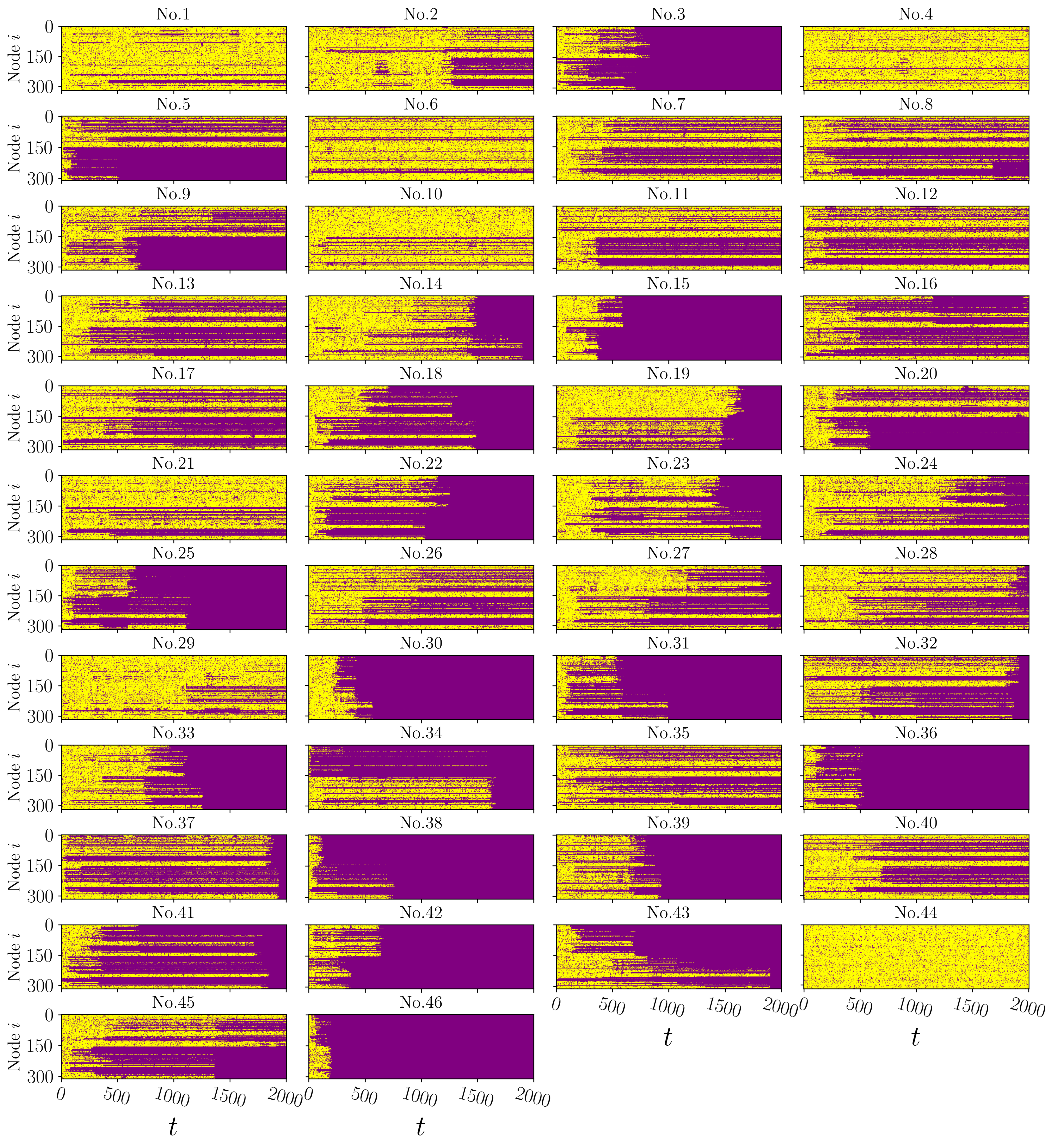}
	\caption{The self-sustaining oscillatory patterns for the individuals in healthy control group. The activation threshold $\omega_{c} = 0.37$.}
	\label{figureS2_con}
\end{figure*}
\begin{figure*}[!h]
	\centering
	\includegraphics[width=1\textwidth]{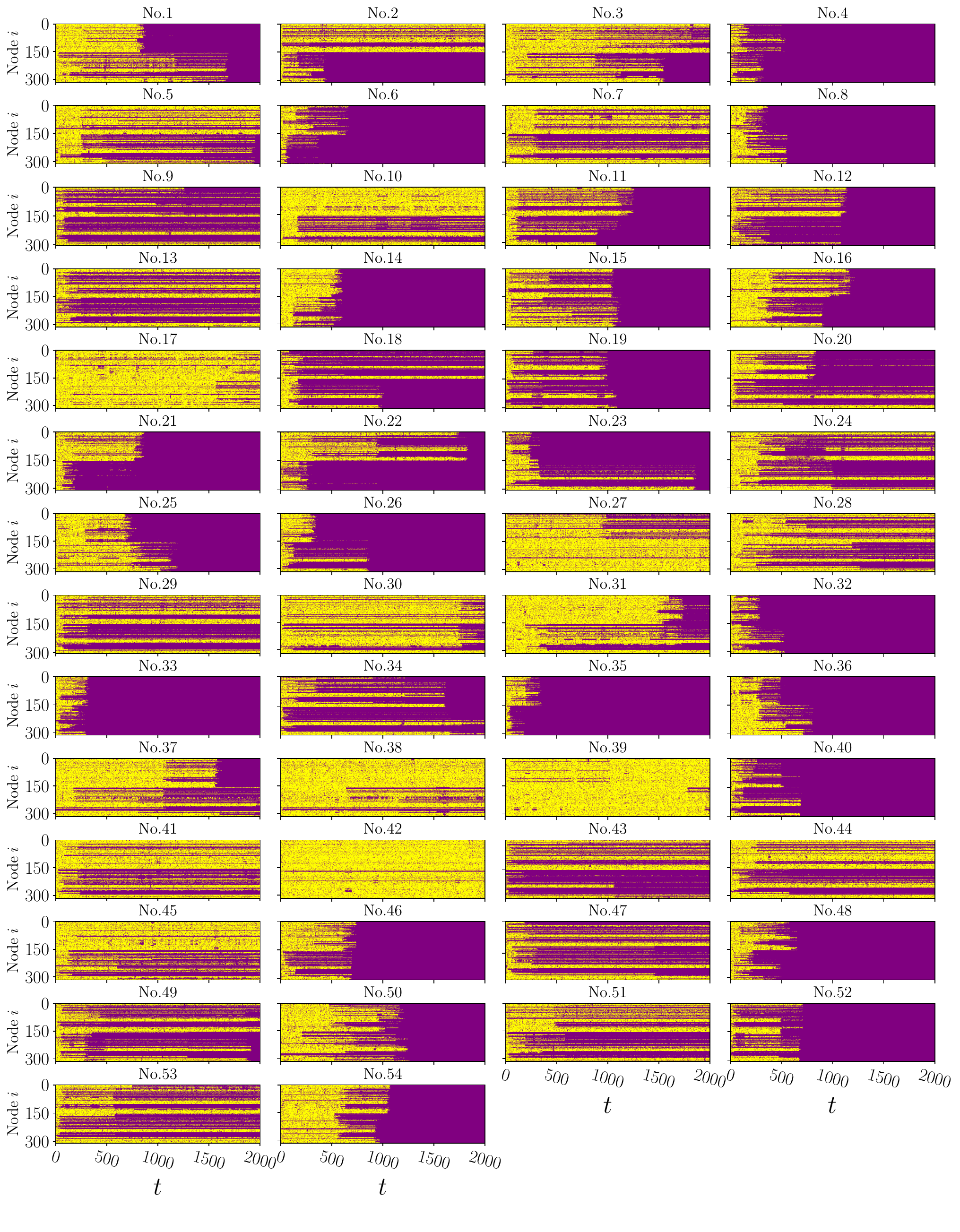}
	\caption{The self-sustaining oscillatory patterns for the individuals in three months post-stoke group. The activation threshold $\omega_{c} = 0.37$.}
	\label{figureS2_pat_t1}
\end{figure*}
\begin{figure*}[!h]
	\centering
	\includegraphics[width=1\textwidth]{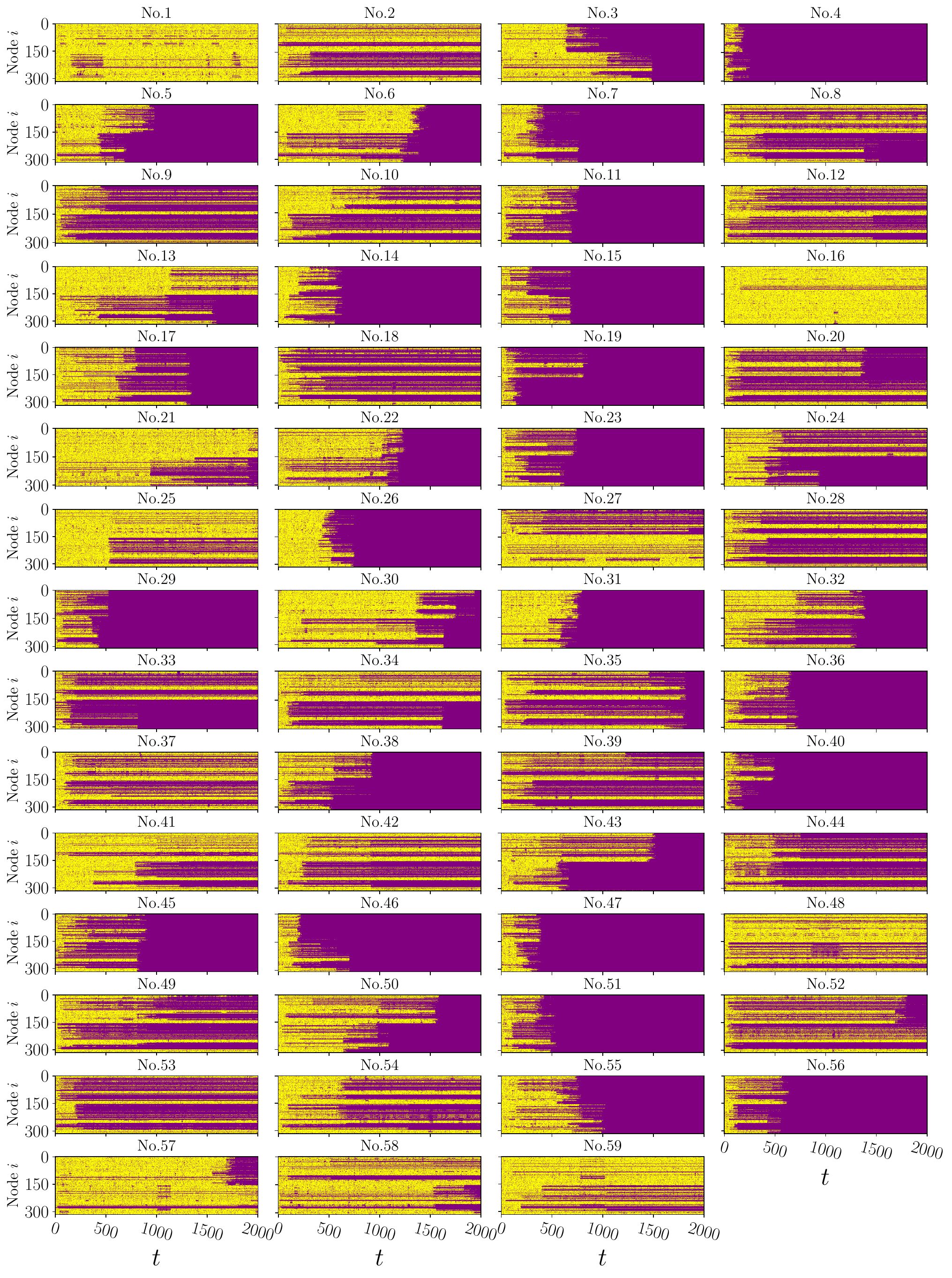}
	\caption{The self-sustaining oscillatory patterns for the individuals in one year post-stoke group. The activation threshold $\omega_{c} = 0.37$.}
	\label{figureS2_pat_t2}
\end{figure*}

We show more results with supplementary figures here to confirm our findings. To verify the approximate solution of the critical $\omega'_c$ in Eq.~($\ref{eq:condition2}$) with a tiny initial seed, we first generate the synthetic fully connected networks with network size $N=1000$. We then assign a constant weight to each edge for a given average weight $\langle w \rangle$. Finally, we simulate the activation propagation model on the synthetic networks and use Eq.~($\ref{eq:8}$) to numerically identify the 
critical $\omega'_c$. All the results have been averaged over $10^3$ independent realizations. We can see that the simulation results agree well with the theoretical analysis in Eq.~($\ref{eq:condition2}$), confirming that the critical $\omega'_c$ depends on the seeds initially selected and the activation threshold.  

We also show the self-sustaining oscillatory patterns for all individuals in Figs.~\ref{figureS2_con}-\ref{figureS2_pat_t2}. We find that the time-limited self-sustaining oscillatory pattern is not unique for the individuals in $Pat.\ t_1$, but also appears in $Con$ and $Pat.\ t_2$. 


\section{Supplementary Tables}
\label{Sup_T}
\begin{table} [!b]
	\centering
	\caption{\textbf{Statistical analysis of network properties across stroke brain networks.} This table presents the p-values and absolute values of Cohen's $|d|$ for comparing global network properties between the control group ($Con$) and two patient groups ($Pat. \ t_1$ and $Pat. \ t_2$). The global properties include the average weight ($\langle w \rangle$), degree ($\langle k \rangle$), maximum degree ($\langle k_{max} \rangle$), clustering coefficient ($\langle C \rangle$), modularity ($\langle Q \rangle$), degree assortativity ($\langle r_c \rangle$), characteristic path length ($\langle L \rangle$), and diameter ($\langle D \rangle$). The p-values indicate the significance of the differences, while Cohen's $|d|$ values measure the effect size. }
	\begin{tabular}{ccccc}
		\toprule
		\multirow{2}{*}{Properties} & \multicolumn{2}{c}{p-value} & \multicolumn{2}{c}{Cohen's $|d|$} \\
		& {Con vs. $t_1$} & {Con vs. $t_2$} & {Con vs. $t_1$} & {Con vs. $t_2$} \\
		\midrule
		\thd{$\langle w\rangle$}        & 0.0001 & 0.0045 & 0.7836 & 0.5551 \\
		\thd{$\langle k\rangle$}        & 0.0004 & 0.0057 & 0.7234 & 0.5591 \\
		\thd{$\langle k_{max}\rangle$}  & 0.0000 & 0.0131 & 0.9135 & 0.4785 \\
		\thd{$\langle C\rangle$}        & 0.0077 & 0.0041 & 0.5311 & 0.5753 \\
		\thd{$\langle Q\rangle$}        & 0.0000 & 0.0001 & 1.0769 & 0.8148 \\
		\thd{$\langle r_c\rangle$}      & 0.0041 & 0.0090 & 0.5797 & 0.4993 \\
		\thd{$\langle L\rangle$}        & 0.0000 & 0.0017 & 0.8609 & 0.5991 \\
		\thd{$\langle D\rangle$}        & 0.0006 & 0.0175 & 0.6792 & 0.4504 \\
		\bottomrule
	\end{tabular}
	\label{table2}
\end{table}

\begin{table}  [!hb]
	\centering
	\caption{\textbf{Statistical analysis of regional brain network differences in spreading dynamics.} This table presents the p-values and absolute values of Cohen's $|d|$ for comparing differences in spreading dynamics across specific brain regions between the control group ($Con$) and two patient groups ($Pat. \ t_1$ and $Pat. \ t_2$). The analyzed regions include the visual network (VIS), somatomotor network (SMD), cingulo-opercular network (CON), dorsal attention network (DAN), frontoparietal network (FPN), default mode network (DMN), and None (NON). The p-values indicate statistical significance, and Cohen's $|d|$ values provide the effect size.}
	\begin{tabular}{ccccc}
		\toprule
		\multirow{2}{*}{Properties} & \multicolumn{2}{c}{p-value} & \multicolumn{2}{c}{Cohen's $|d|$} \\
		& {Con vs. $t_1$} & {Con vs. $t_2$} & {Con vs. $t_1$} & {Con vs. $t_2$} \\
		\midrule
		\thd{VIS} & 0.0000 & 0.0193 & 0.8637 & 0.4482 \\
		\thd{SMD} & 0.0000 & 0.0090 & 0.8399 & 0.4870 \\
		\thd{CON} & 0.0046 & 0.0139 & 0.5619 & 0.4666 \\
		\thd{DAN} & 0.0000 & 0.0195 & 0.8675 & 0.4502 \\
		\thd{FPN} & 0.0033 & 0.1224 & 0.5819 & 0.2987 \\
		\thd{DMN} & 0.0723 & 0.6512 & 0.3525 & 0.0891 \\
		\thd{NON} & 0.0003 & 0.0103 & 0.7237 & 0.4903 \\
		\bottomrule
	\end{tabular}
	\label{table3}
\end{table}

\section*{Acknowledgments}
This work was supported by National Natural
Science Foundation of China (Grants No.~12305043, No.~12165016 and No.~12005079), the Natural Science Foundation of Jiangsu Province (Grant No.~BK20220511), 
the funding for Scientific Research Startup of Jiangsu University (Grant No.~4111710001 and No.~4111190017), 
the Project of Undergraduate Scientific Research (Grant No.~22A418). M. Z. appreciates the support from the Jiangsu Specially-Appointed Professor Program.



\begin{thebibliography}{62}%
\makeatletter
\providecommand \@ifxundefined [1]{%
 \@ifx{#1\undefined}
}%
\providecommand \@ifnum [1]{%
 \ifnum #1\expandafter \@firstoftwo
 \else \expandafter \@secondoftwo
 \fi
}%
\providecommand \@ifx [1]{%
 \ifx #1\expandafter \@firstoftwo
 \else \expandafter \@secondoftwo
 \fi
}%
\providecommand \natexlab [1]{#1}%
\providecommand \enquote  [1]{``#1''}%
\providecommand \bibnamefont  [1]{#1}%
\providecommand \bibfnamefont [1]{#1}%
\providecommand \citenamefont [1]{#1}%
\providecommand \href@noop [0]{\@secondoftwo}%
\providecommand \href [0]{\begingroup \@sanitize@url \@href}%
\providecommand \@href[1]{\@@startlink{#1}\@@href}%
\providecommand \@@href[1]{\endgroup#1\@@endlink}%
\providecommand \@sanitize@url [0]{\catcode `\\12\catcode `\$12\catcode
  `\&12\catcode `\#12\catcode `\^12\catcode `\_12\catcode `\%12\relax}%
\providecommand \@@startlink[1]{}%
\providecommand \@@endlink[0]{}%
\providecommand \url  [0]{\begingroup\@sanitize@url \@url }%
\providecommand \@url [1]{\endgroup\@href {#1}{\urlprefix }}%
\providecommand \urlprefix  [0]{URL }%
\providecommand \Eprint [0]{\href }%
\providecommand \doibase [0]{https://doi.org/}%
\providecommand \selectlanguage [0]{\@gobble}%
\providecommand \bibinfo  [0]{\@secondoftwo}%
\providecommand \bibfield  [0]{\@secondoftwo}%
\providecommand \translation [1]{[#1]}%
\providecommand \BibitemOpen [0]{}%
\providecommand \bibitemStop [0]{}%
\providecommand \bibitemNoStop [0]{.\EOS\space}%
\providecommand \EOS [0]{\spacefactor3000\relax}%
\providecommand \BibitemShut  [1]{\csname bibitem#1\endcsname}%
\let\auto@bib@innerbib\@empty
\bibitem [{\citenamefont {Bullmore}\ and\ \citenamefont
  {Sporns}(2009)}]{Bullmore2009}%
  \BibitemOpen
  \bibfield  {author} {\bibinfo {author} {\bibfnamefont {E.}~\bibnamefont
  {Bullmore}}\ and\ \bibinfo {author} {\bibfnamefont {O.}~\bibnamefont
  {Sporns}},\ }\bibfield  {title} {\bibinfo {title} {Complex brain networks:
  graph theoretical analysis of structural and functional systems},\ }\href
  {https://www.nature.com/articles/nrn2575} {\bibfield  {journal} {\bibinfo
  {journal} {Nature Reviews Neuroscience}\ }\textbf {\bibinfo {volume} {10}},\
  \bibinfo {pages} {186} (\bibinfo {year} {2009})}\BibitemShut {NoStop}%
\bibitem [{\citenamefont {Rubinov}\ and\ \citenamefont
  {Sporns}(2010)}]{Rubinov2010}%
  \BibitemOpen
  \bibfield  {author} {\bibinfo {author} {\bibfnamefont {M.}~\bibnamefont
  {Rubinov}}\ and\ \bibinfo {author} {\bibfnamefont {O.}~\bibnamefont
  {Sporns}},\ }\bibfield  {title} {\bibinfo {title} {Complex network measures
  of brain connectivity: uses and interpretations},\ }\href
  {https://www.sciencedirect.com/science/article/abs/pii/S105381190901074X}
  {\bibfield  {journal} {\bibinfo  {journal} {NeuroImage}\ }\textbf {\bibinfo
  {volume} {52}},\ \bibinfo {pages} {1059} (\bibinfo {year}
  {2010})}\BibitemShut {NoStop}%
\bibitem [{\citenamefont {Fornito}\ \emph {et~al.}(2016)\citenamefont
  {Fornito}, \citenamefont {Zalesky},\ and\ \citenamefont
  {Bullmore}}]{Fornito2016}%
  \BibitemOpen
  \bibfield  {author} {\bibinfo {author} {\bibfnamefont {A.}~\bibnamefont
  {Fornito}}, \bibinfo {author} {\bibfnamefont {A.}~\bibnamefont {Zalesky}},\
  and\ \bibinfo {author} {\bibfnamefont {E.}~\bibnamefont {Bullmore}},\
  }\href@noop {} {\emph {\bibinfo {title} {Fundamentals of brain network
  analysis}}}\ (\bibinfo  {publisher} {Academic Press},\ \bibinfo {year}
  {2016})\BibitemShut {NoStop}%
\bibitem [{\citenamefont {Watts}\ and\ \citenamefont
  {Strogatz}(1998)}]{Watts1998}%
  \BibitemOpen
  \bibfield  {author} {\bibinfo {author} {\bibfnamefont {D.~J.}\ \bibnamefont
  {Watts}}\ and\ \bibinfo {author} {\bibfnamefont {S.~H.}\ \bibnamefont
  {Strogatz}},\ }\bibfield  {title} {\bibinfo {title} {Collective dynamics of
  ‘small-world’ networks},\ }\href {https://doi.org/10.1038/30918}
  {\bibfield  {journal} {\bibinfo  {journal} {Nature}\ }\textbf {\bibinfo
  {volume} {393}},\ \bibinfo {pages} {440} (\bibinfo {year}
  {1998})}\BibitemShut {NoStop}%
\bibitem [{\citenamefont {Sporns}\ \emph {et~al.}(2004)\citenamefont {Sporns},
  \citenamefont {Chialvo}, \citenamefont {Kaiser},\ and\ \citenamefont
  {Hilgetag}}]{Sporns2004}%
  \BibitemOpen
  \bibfield  {author} {\bibinfo {author} {\bibfnamefont {O.}~\bibnamefont
  {Sporns}}, \bibinfo {author} {\bibfnamefont {D.~R.}\ \bibnamefont {Chialvo}},
  \bibinfo {author} {\bibfnamefont {M.}~\bibnamefont {Kaiser}},\ and\ \bibinfo
  {author} {\bibfnamefont {C.~C.}\ \bibnamefont {Hilgetag}},\ }\bibfield
  {title} {\bibinfo {title} {Organization, development and function of complex
  brain networks},\ }\href
  {https://www.cell.com/ajhg/abstract/S1364-6613(04)00190-1} {\bibfield
  {journal} {\bibinfo  {journal} {Trends in Cognitive Sciences}\ }\textbf
  {\bibinfo {volume} {8}},\ \bibinfo {pages} {418} (\bibinfo {year}
  {2004})}\BibitemShut {NoStop}%
\bibitem [{\citenamefont {Hagmann}\ \emph {et~al.}(2007)\citenamefont
  {Hagmann}, \citenamefont {Kurant}, \citenamefont {Gigandet}, \citenamefont
  {Thiran}, \citenamefont {Wedeen}, \citenamefont {Meuli},\ and\ \citenamefont
  {Thiran}}]{Hagmann2007}%
  \BibitemOpen
  \bibfield  {author} {\bibinfo {author} {\bibfnamefont {P.}~\bibnamefont
  {Hagmann}}, \bibinfo {author} {\bibfnamefont {M.}~\bibnamefont {Kurant}},
  \bibinfo {author} {\bibfnamefont {X.}~\bibnamefont {Gigandet}}, \bibinfo
  {author} {\bibfnamefont {P.}~\bibnamefont {Thiran}}, \bibinfo {author}
  {\bibfnamefont {V.~J.}\ \bibnamefont {Wedeen}}, \bibinfo {author}
  {\bibfnamefont {R.}~\bibnamefont {Meuli}},\ and\ \bibinfo {author}
  {\bibfnamefont {J.-P.}\ \bibnamefont {Thiran}},\ }\bibfield  {title}
  {\bibinfo {title} {Mapping human whole-brain structural networks with
  diffusion mri},\ }\href
  {https://journals.plos.org/plosone/article?id=10.1371/journal.pone.0000597}
  {\bibfield  {journal} {\bibinfo  {journal} {PLOS ONE}\ }\textbf {\bibinfo
  {volume} {2}},\ \bibinfo {pages} {e597} (\bibinfo {year} {2007})}\BibitemShut
  {NoStop}%
\bibitem [{\citenamefont {Gong}\ \emph {et~al.}(2008)\citenamefont {Gong},
  \citenamefont {He}, \citenamefont {Concha}, \citenamefont {Lebel},
  \citenamefont {Gross}, \citenamefont {Evans},\ and\ \citenamefont
  {Beaulieu}}]{Gong2008}%
  \BibitemOpen
  \bibfield  {author} {\bibinfo {author} {\bibfnamefont {G.}~\bibnamefont
  {Gong}}, \bibinfo {author} {\bibfnamefont {Y.}~\bibnamefont {He}}, \bibinfo
  {author} {\bibfnamefont {L.}~\bibnamefont {Concha}}, \bibinfo {author}
  {\bibfnamefont {C.}~\bibnamefont {Lebel}}, \bibinfo {author} {\bibfnamefont
  {D.~W.}\ \bibnamefont {Gross}}, \bibinfo {author} {\bibfnamefont {A.~C.}\
  \bibnamefont {Evans}},\ and\ \bibinfo {author} {\bibfnamefont
  {C.}~\bibnamefont {Beaulieu}},\ }\bibfield  {title} {\bibinfo {title}
  {Mapping anatomical connectivity patterns of human cerebral cortex using in
  vivo diffusion tensor imaging tractography},\ }\href
  {https://academic.oup.com/cercor/article-abstract/19/3/524/428176} {\bibfield
   {journal} {\bibinfo  {journal} {Cerebral Cortex}\ }\textbf {\bibinfo
  {volume} {19}},\ \bibinfo {pages} {524} (\bibinfo {year} {2008})}\BibitemShut
  {NoStop}%
\bibitem [{\citenamefont {Crossley}\ \emph {et~al.}(2014)\citenamefont
  {Crossley}, \citenamefont {Mechelli}, \citenamefont {Scott}, \citenamefont
  {Carletti}, \citenamefont {Fox}, \citenamefont {McGuire},\ and\ \citenamefont
  {Bullmore}}]{Crossley2014}%
  \BibitemOpen
  \bibfield  {author} {\bibinfo {author} {\bibfnamefont {N.~A.}\ \bibnamefont
  {Crossley}}, \bibinfo {author} {\bibfnamefont {A.}~\bibnamefont {Mechelli}},
  \bibinfo {author} {\bibfnamefont {J.}~\bibnamefont {Scott}}, \bibinfo
  {author} {\bibfnamefont {F.}~\bibnamefont {Carletti}}, \bibinfo {author}
  {\bibfnamefont {P.~T.}\ \bibnamefont {Fox}}, \bibinfo {author} {\bibfnamefont
  {P.}~\bibnamefont {McGuire}},\ and\ \bibinfo {author} {\bibfnamefont {E.~T.}\
  \bibnamefont {Bullmore}},\ }\bibfield  {title} {\bibinfo {title} {The hubs of
  the human connectome are generally implicated in the anatomy of brain
  disorders},\ }\href
  {https://academic.oup.com/brain/article/137/8/2382/2847927} {\bibfield
  {journal} {\bibinfo  {journal} {Brain}\ }\textbf {\bibinfo {volume} {137}},\
  \bibinfo {pages} {2382} (\bibinfo {year} {2014})}\BibitemShut {NoStop}%
\bibitem [{\citenamefont {van~den Heuvel}\ and\ \citenamefont
  {Sporns}(2011)}]{VandenHeuvel2011}%
  \BibitemOpen
  \bibfield  {author} {\bibinfo {author} {\bibfnamefont {M.~P.}\ \bibnamefont
  {van~den Heuvel}}\ and\ \bibinfo {author} {\bibfnamefont {O.}~\bibnamefont
  {Sporns}},\ }\bibfield  {title} {\bibinfo {title} {{Rich-club organization of
  the human connectome.}},\ }\href
  {https://doi.org/10.1523/JNEUROSCI.3539-11.2011} {\bibfield  {journal}
  {\bibinfo  {journal} {Journal of Neuroscience}\ }\textbf {\bibinfo {volume}
  {31}},\ \bibinfo {pages} {15775} (\bibinfo {year} {2011})}\BibitemShut
  {NoStop}%
\bibitem [{\citenamefont {Sporns}\ and\ \citenamefont
  {Betzel}(2016)}]{Sporns2016}%
  \BibitemOpen
  \bibfield  {author} {\bibinfo {author} {\bibfnamefont {O.}~\bibnamefont
  {Sporns}}\ and\ \bibinfo {author} {\bibfnamefont {R.~F.}\ \bibnamefont
  {Betzel}},\ }\bibfield  {title} {\bibinfo {title} {Modular brain networks},\
  }\href
  {https://www.annualreviews.org/content/journals/10.1146/annurev-psych-122414-033634}
  {\bibfield  {journal} {\bibinfo  {journal} {Annual Review of Psychology}\
  }\textbf {\bibinfo {volume} {67}},\ \bibinfo {pages} {613} (\bibinfo {year}
  {2016})}\BibitemShut {NoStop}%
\bibitem [{\citenamefont {Meunier}\ \emph
  {et~al.}(2010{\natexlab{a}})\citenamefont {Meunier}, \citenamefont
  {Lambiotte},\ and\ \citenamefont {Bullmore}}]{Meunier2010}%
  \BibitemOpen
  \bibfield  {author} {\bibinfo {author} {\bibfnamefont {D.}~\bibnamefont
  {Meunier}}, \bibinfo {author} {\bibfnamefont {R.}~\bibnamefont {Lambiotte}},\
  and\ \bibinfo {author} {\bibfnamefont {E.~T.}\ \bibnamefont {Bullmore}},\
  }\bibfield  {title} {\bibinfo {title} {{Modular and Hierarchically Modular
  Organization of Brain Networks}},\ }\href
  {https://doi.org/10.3389/fnins.2010.00200} {\bibfield  {journal} {\bibinfo
  {journal} {Frontiers in Neuroscience}\ }\textbf {\bibinfo {volume} {4}},\
  \bibinfo {pages} {200} (\bibinfo {year} {2010}{\natexlab{a}})}\BibitemShut
  {NoStop}%
\bibitem [{\citenamefont {Avena-Koenigsberger}\ \emph
  {et~al.}(2018)\citenamefont {Avena-Koenigsberger}, \citenamefont {Misic},\
  and\ \citenamefont {Sporns}}]{avena2018communication}%
  \BibitemOpen
  \bibfield  {author} {\bibinfo {author} {\bibfnamefont {A.}~\bibnamefont
  {Avena-Koenigsberger}}, \bibinfo {author} {\bibfnamefont {B.}~\bibnamefont
  {Misic}},\ and\ \bibinfo {author} {\bibfnamefont {O.}~\bibnamefont
  {Sporns}},\ }\bibfield  {title} {\bibinfo {title} {Communication dynamics in
  complex brain networks},\ }\href
  {https://www.nature.com/articles/nrn.2017.149} {\bibfield  {journal}
  {\bibinfo  {journal} {Nature Reviews Neuroscience}\ }\textbf {\bibinfo
  {volume} {19}},\ \bibinfo {pages} {17} (\bibinfo {year} {2018})}\BibitemShut
  {NoStop}%
\bibitem [{\citenamefont {Bullmore}\ and\ \citenamefont
  {Sporns}(2012)}]{bullmore2012economy}%
  \BibitemOpen
  \bibfield  {author} {\bibinfo {author} {\bibfnamefont {E.}~\bibnamefont
  {Bullmore}}\ and\ \bibinfo {author} {\bibfnamefont {O.}~\bibnamefont
  {Sporns}},\ }\bibfield  {title} {\bibinfo {title} {The economy of brain
  network organization},\ }\href {https://www.nature.com/articles/nrn3214}
  {\bibfield  {journal} {\bibinfo  {journal} {Nature Reviews Neuroscience}\
  }\textbf {\bibinfo {volume} {13}},\ \bibinfo {pages} {336} (\bibinfo {year}
  {2012})}\BibitemShut {NoStop}%
\bibitem [{\citenamefont {Zheng}\ \emph {et~al.}(2020)\citenamefont {Zheng},
  \citenamefont {Allard}, \citenamefont {Hagmann}, \citenamefont
  {Alem{\'a}n-G{\'o}mez},\ and\ \citenamefont {Serrano}}]{zheng2020geometric}%
  \BibitemOpen
  \bibfield  {author} {\bibinfo {author} {\bibfnamefont {M.}~\bibnamefont
  {Zheng}}, \bibinfo {author} {\bibfnamefont {A.}~\bibnamefont {Allard}},
  \bibinfo {author} {\bibfnamefont {P.}~\bibnamefont {Hagmann}}, \bibinfo
  {author} {\bibfnamefont {Y.}~\bibnamefont {Alem{\'a}n-G{\'o}mez}},\ and\
  \bibinfo {author} {\bibfnamefont {M.~{\'A}.}\ \bibnamefont {Serrano}},\
  }\bibfield  {title} {\bibinfo {title} {Geometric renormalization unravels
  self-similarity of the multiscale human connectome},\ }\href
  {https://www.pnas.org/doi/abs/10.1073/pnas.1922248117} {\bibfield  {journal}
  {\bibinfo  {journal} {Proceedings of the National Academy of Sciences}\
  }\textbf {\bibinfo {volume} {117}},\ \bibinfo {pages} {20244} (\bibinfo
  {year} {2020})}\BibitemShut {NoStop}%
\bibitem [{\citenamefont {Tononi}(2010)}]{tononi2010information}%
  \BibitemOpen
  \bibfield  {author} {\bibinfo {author} {\bibfnamefont {G.}~\bibnamefont
  {Tononi}},\ }\bibfield  {title} {\bibinfo {title} {Information integration:
  its relevance to brain function and consciousness},\ }\href
  {https://pubmed.ncbi.nlm.nih.gov/21175016/} {\bibfield  {journal} {\bibinfo
  {journal} {Archives Italiennes de Biologie}\ }\textbf {\bibinfo {volume}
  {148}},\ \bibinfo {pages} {299} (\bibinfo {year} {2010})}\BibitemShut
  {NoStop}%
\bibitem [{\citenamefont {Sporns}(2013)}]{sporns2013network}%
  \BibitemOpen
  \bibfield  {author} {\bibinfo {author} {\bibfnamefont {O.}~\bibnamefont
  {Sporns}},\ }\bibfield  {title} {\bibinfo {title} {Network attributes for
  segregation and integration in the human brain},\ }\href
  {https://www.sciencedirect.com/science/article/abs/pii/S0959438812001894}
  {\bibfield  {journal} {\bibinfo  {journal} {Current Opinion in Neurobiology}\
  }\textbf {\bibinfo {volume} {23}},\ \bibinfo {pages} {162} (\bibinfo {year}
  {2013})}\BibitemShut {NoStop}%
\bibitem [{\citenamefont {Hilgetag}\ and\ \citenamefont
  {Goulas}(2020)}]{hilgetag2020hierarchy}%
  \BibitemOpen
  \bibfield  {author} {\bibinfo {author} {\bibfnamefont {C.~C.}\ \bibnamefont
  {Hilgetag}}\ and\ \bibinfo {author} {\bibfnamefont {A.}~\bibnamefont
  {Goulas}},\ }\bibfield  {title} {\bibinfo {title} {‘hierarchy’in the
  organization of brain networks},\ }\href
  {https://royalsocietypublishing.org/doi/full/10.1098/rstb.2019.0319}
  {\bibfield  {journal} {\bibinfo  {journal} {Philosophical Transactions of the
  Royal Society B}\ }\textbf {\bibinfo {volume} {375}},\ \bibinfo {pages}
  {20190319} (\bibinfo {year} {2020})}\BibitemShut {NoStop}%
\bibitem [{\citenamefont {Meunier}\ \emph
  {et~al.}(2010{\natexlab{b}})\citenamefont {Meunier}, \citenamefont
  {Lambiotte},\ and\ \citenamefont {Bullmore}}]{meunier2010modular}%
  \BibitemOpen
  \bibfield  {author} {\bibinfo {author} {\bibfnamefont {D.}~\bibnamefont
  {Meunier}}, \bibinfo {author} {\bibfnamefont {R.}~\bibnamefont {Lambiotte}},\
  and\ \bibinfo {author} {\bibfnamefont {E.~T.}\ \bibnamefont {Bullmore}},\
  }\bibfield  {title} {\bibinfo {title} {Modular and hierarchically modular
  organization of brain networks},\ }\href
  {https://www.frontiersin.org/journals/neuroscience/articles/10.3389/fnins.2010.00200/full}
  {\bibfield  {journal} {\bibinfo  {journal} {Frontiers in Neuroscience}\
  }\textbf {\bibinfo {volume} {4}},\ \bibinfo {pages} {200} (\bibinfo {year}
  {2010}{\natexlab{b}})}\BibitemShut {NoStop}%
\bibitem [{\citenamefont {Mi{\v{s}}i{\'c}}\ \emph {et~al.}(2015)\citenamefont
  {Mi{\v{s}}i{\'c}}, \citenamefont {Betzel}, \citenamefont {Nematzadeh},
  \citenamefont {Goni}, \citenamefont {Griffa}, \citenamefont {Hagmann},
  \citenamefont {Flammini}, \citenamefont {Ahn},\ and\ \citenamefont
  {Sporns}}]{mivsic2015cooperative}%
  \BibitemOpen
  \bibfield  {author} {\bibinfo {author} {\bibfnamefont {B.}~\bibnamefont
  {Mi{\v{s}}i{\'c}}}, \bibinfo {author} {\bibfnamefont {R.~F.}\ \bibnamefont
  {Betzel}}, \bibinfo {author} {\bibfnamefont {A.}~\bibnamefont {Nematzadeh}},
  \bibinfo {author} {\bibfnamefont {J.}~\bibnamefont {Goni}}, \bibinfo {author}
  {\bibfnamefont {A.}~\bibnamefont {Griffa}}, \bibinfo {author} {\bibfnamefont
  {P.}~\bibnamefont {Hagmann}}, \bibinfo {author} {\bibfnamefont
  {A.}~\bibnamefont {Flammini}}, \bibinfo {author} {\bibfnamefont {Y.-Y.}\
  \bibnamefont {Ahn}},\ and\ \bibinfo {author} {\bibfnamefont {O.}~\bibnamefont
  {Sporns}},\ }\bibfield  {title} {\bibinfo {title} {Cooperative and
  competitive spreading dynamics on the human connectome},\ }\href
  {https://www.cell.com/neuron/fulltext/S0896-6273(15)00474-2} {\bibfield
  {journal} {\bibinfo  {journal} {Neuron}\ }\textbf {\bibinfo {volume} {86}},\
  \bibinfo {pages} {1518} (\bibinfo {year} {2015})}\BibitemShut {NoStop}%
\bibitem [{\citenamefont {Carr}\ \emph {et~al.}(1993)\citenamefont {Carr} \emph
  {et~al.}}]{carr1993processing}%
  \BibitemOpen
  \bibfield  {author} {\bibinfo {author} {\bibfnamefont {C.~E.}\ \bibnamefont
  {Carr}} \emph {et~al.},\ }\bibfield  {title} {\bibinfo {title} {Processing of
  temporal information in the brain},\ }\href
  {https://www.annualreviews.org/content/journals/10.1146/annurev.ne.16.030193.001255}
  {\bibfield  {journal} {\bibinfo  {journal} {Annual Review of Neuroscience}\
  }\textbf {\bibinfo {volume} {16}},\ \bibinfo {pages} {223} (\bibinfo {year}
  {1993})}\BibitemShut {NoStop}%
\bibitem [{\citenamefont {Buonomano}\ and\ \citenamefont
  {Maass}(2009)}]{buonomano2009state}%
  \BibitemOpen
  \bibfield  {author} {\bibinfo {author} {\bibfnamefont {D.~V.}\ \bibnamefont
  {Buonomano}}\ and\ \bibinfo {author} {\bibfnamefont {W.}~\bibnamefont
  {Maass}},\ }\bibfield  {title} {\bibinfo {title} {State-dependent
  computations: spatiotemporal processing in cortical networks},\ }\href
  {https://www.nature.com/articles/nrn2558} {\bibfield  {journal} {\bibinfo
  {journal} {Nature Reviews Neuroscience}\ }\textbf {\bibinfo {volume} {10}},\
  \bibinfo {pages} {113} (\bibinfo {year} {2009})}\BibitemShut {NoStop}%
\bibitem [{\citenamefont {Mongillo}\ \emph {et~al.}(2008)\citenamefont
  {Mongillo}, \citenamefont {Barak},\ and\ \citenamefont
  {Tsodyks}}]{mongillo2008synaptic}%
  \BibitemOpen
  \bibfield  {author} {\bibinfo {author} {\bibfnamefont {G.}~\bibnamefont
  {Mongillo}}, \bibinfo {author} {\bibfnamefont {O.}~\bibnamefont {Barak}},\
  and\ \bibinfo {author} {\bibfnamefont {M.}~\bibnamefont {Tsodyks}},\
  }\bibfield  {title} {\bibinfo {title} {Synaptic theory of working memory},\
  }\href {https://www.science.org/doi/10.1126/science.1150769} {\bibfield
  {journal} {\bibinfo  {journal} {Science}\ }\textbf {\bibinfo {volume}
  {319}},\ \bibinfo {pages} {1543} (\bibinfo {year} {2008})}\BibitemShut
  {NoStop}%
\bibitem [{\citenamefont {Mi}\ \emph {et~al.}(2013)\citenamefont {Mi},
  \citenamefont {Liao}, \citenamefont {Huang}, \citenamefont {Zhang},
  \citenamefont {Gu}, \citenamefont {Hu},\ and\ \citenamefont
  {Wu}}]{mi2013long}%
  \BibitemOpen
  \bibfield  {author} {\bibinfo {author} {\bibfnamefont {Y.}~\bibnamefont
  {Mi}}, \bibinfo {author} {\bibfnamefont {X.}~\bibnamefont {Liao}}, \bibinfo
  {author} {\bibfnamefont {X.}~\bibnamefont {Huang}}, \bibinfo {author}
  {\bibfnamefont {L.}~\bibnamefont {Zhang}}, \bibinfo {author} {\bibfnamefont
  {W.}~\bibnamefont {Gu}}, \bibinfo {author} {\bibfnamefont {G.}~\bibnamefont
  {Hu}},\ and\ \bibinfo {author} {\bibfnamefont {S.}~\bibnamefont {Wu}},\
  }\bibfield  {title} {\bibinfo {title} {Long-period rhythmic synchronous
  firing in a scale-free network},\ }\href
  {https://www.pnas.org/doi/abs/10.1073/pnas.1304680110} {\bibfield  {journal}
  {\bibinfo  {journal} {Proceedings of the National Academy of Sciences}\
  }\textbf {\bibinfo {volume} {110}},\ \bibinfo {pages} {E4931} (\bibinfo
  {year} {2013})}\BibitemShut {NoStop}%
\bibitem [{\citenamefont {Xu}\ \emph {et~al.}(2014)\citenamefont {Xu},
  \citenamefont {Zhang}, \citenamefont {Wang},\ and\ \citenamefont
  {Liu}}]{xu2014simplified}%
  \BibitemOpen
  \bibfield  {author} {\bibinfo {author} {\bibfnamefont {K.}~\bibnamefont
  {Xu}}, \bibinfo {author} {\bibfnamefont {X.}~\bibnamefont {Zhang}}, \bibinfo
  {author} {\bibfnamefont {C.}~\bibnamefont {Wang}},\ and\ \bibinfo {author}
  {\bibfnamefont {Z.}~\bibnamefont {Liu}},\ }\bibfield  {title} {\bibinfo
  {title} {A simplified memory network model based on pattern formations},\
  }\href {https://www.nature.com/articles/srep07568} {\bibfield  {journal}
  {\bibinfo  {journal} {Scientific Reports}\ }\textbf {\bibinfo {volume} {4}},\
  \bibinfo {pages} {7568} (\bibinfo {year} {2014})}\BibitemShut {NoStop}%
\bibitem [{\citenamefont {Xu}\ \emph {et~al.}(2013)\citenamefont {Xu},
  \citenamefont {Huang}, \citenamefont {Li}, \citenamefont {Dhamala},\ and\
  \citenamefont {Liu}}]{xu2013controlling}%
  \BibitemOpen
  \bibfield  {author} {\bibinfo {author} {\bibfnamefont {K.}~\bibnamefont
  {Xu}}, \bibinfo {author} {\bibfnamefont {W.}~\bibnamefont {Huang}}, \bibinfo
  {author} {\bibfnamefont {B.}~\bibnamefont {Li}}, \bibinfo {author}
  {\bibfnamefont {M.}~\bibnamefont {Dhamala}},\ and\ \bibinfo {author}
  {\bibfnamefont {Z.}~\bibnamefont {Liu}},\ }\bibfield  {title} {\bibinfo
  {title} {Controlling self-sustained spiking activity by adding or removing
  one network link},\ }\href
  {https://iopscience.iop.org/article/10.1209/0295-5075/102/50002/meta}
  {\bibfield  {journal} {\bibinfo  {journal} {Europhysics Letters}\ }\textbf
  {\bibinfo {volume} {102}},\ \bibinfo {pages} {50002} (\bibinfo {year}
  {2013})}\BibitemShut {NoStop}%
\bibitem [{\citenamefont {Roxin}\ \emph {et~al.}(2004)\citenamefont {Roxin},
  \citenamefont {Riecke},\ and\ \citenamefont {Solla}}]{roxin2004self}%
  \BibitemOpen
  \bibfield  {author} {\bibinfo {author} {\bibfnamefont {A.}~\bibnamefont
  {Roxin}}, \bibinfo {author} {\bibfnamefont {H.}~\bibnamefont {Riecke}},\ and\
  \bibinfo {author} {\bibfnamefont {S.~A.}\ \bibnamefont {Solla}},\ }\bibfield
  {title} {\bibinfo {title} {Self-sustained activity in a small-world network
  of excitable neurons},\ }\href@noop {} {\bibfield  {journal} {\bibinfo
  {journal} {Physical Review Letters}\ }\textbf {\bibinfo {volume} {92}},\
  \bibinfo {pages} {198101} (\bibinfo {year} {2004})}\BibitemShut {NoStop}%
\bibitem [{\citenamefont {Bansal}\ \emph {et~al.}(2019)\citenamefont {Bansal},
  \citenamefont {Garcia}, \citenamefont {Tompson}, \citenamefont {Verstynen},
  \citenamefont {Vettel},\ and\ \citenamefont {Muldoon}}]{bansal2019cognitive}%
  \BibitemOpen
  \bibfield  {author} {\bibinfo {author} {\bibfnamefont {K.}~\bibnamefont
  {Bansal}}, \bibinfo {author} {\bibfnamefont {J.~O.}\ \bibnamefont {Garcia}},
  \bibinfo {author} {\bibfnamefont {S.~H.}\ \bibnamefont {Tompson}}, \bibinfo
  {author} {\bibfnamefont {T.}~\bibnamefont {Verstynen}}, \bibinfo {author}
  {\bibfnamefont {J.~M.}\ \bibnamefont {Vettel}},\ and\ \bibinfo {author}
  {\bibfnamefont {S.~F.}\ \bibnamefont {Muldoon}},\ }\bibfield  {title}
  {\bibinfo {title} {Cognitive chimera states in human brain networks},\ }\href
  {https://www.science.org/doi/10.1126/sciadv.aau8535} {\bibfield  {journal}
  {\bibinfo  {journal} {Science Advances}\ }\textbf {\bibinfo {volume} {5}},\
  \bibinfo {pages} {eaau8535} (\bibinfo {year} {2019})}\BibitemShut {NoStop}%
\bibitem [{\citenamefont {Honey}\ \emph {et~al.}(2009)\citenamefont {Honey},
  \citenamefont {Sporns}, \citenamefont {Cammoun}, \citenamefont {Gigandet},
  \citenamefont {Thiran}, \citenamefont {Meuli},\ and\ \citenamefont
  {Hagmann}}]{honey2009predicting}%
  \BibitemOpen
  \bibfield  {author} {\bibinfo {author} {\bibfnamefont {C.~J.}\ \bibnamefont
  {Honey}}, \bibinfo {author} {\bibfnamefont {O.}~\bibnamefont {Sporns}},
  \bibinfo {author} {\bibfnamefont {L.}~\bibnamefont {Cammoun}}, \bibinfo
  {author} {\bibfnamefont {X.}~\bibnamefont {Gigandet}}, \bibinfo {author}
  {\bibfnamefont {J.-P.}\ \bibnamefont {Thiran}}, \bibinfo {author}
  {\bibfnamefont {R.}~\bibnamefont {Meuli}},\ and\ \bibinfo {author}
  {\bibfnamefont {P.}~\bibnamefont {Hagmann}},\ }\bibfield  {title} {\bibinfo
  {title} {Predicting human resting-state functional connectivity from
  structural connectivity},\ }\href
  {https://www.pnas.org/doi/full/10.1073/pnas.0811168106} {\bibfield  {journal}
  {\bibinfo  {journal} {Proceedings of the National Academy of Sciences}\
  }\textbf {\bibinfo {volume} {106}},\ \bibinfo {pages} {2035} (\bibinfo {year}
  {2009})}\BibitemShut {NoStop}%
\bibitem [{\citenamefont {Kinouchi}\ and\ \citenamefont
  {Copelli}(2006)}]{kinouchi2006optimal}%
  \BibitemOpen
  \bibfield  {author} {\bibinfo {author} {\bibfnamefont {O.}~\bibnamefont
  {Kinouchi}}\ and\ \bibinfo {author} {\bibfnamefont {M.}~\bibnamefont
  {Copelli}},\ }\bibfield  {title} {\bibinfo {title} {Optimal dynamical range
  of excitable networks at criticality},\ }\href
  {https://www.nature.com/articles/nphys289} {\bibfield  {journal} {\bibinfo
  {journal} {Nature Physics}\ }\textbf {\bibinfo {volume} {2}},\ \bibinfo
  {pages} {348} (\bibinfo {year} {2006})}\BibitemShut {NoStop}%
\bibitem [{\citenamefont {Brunel}\ and\ \citenamefont
  {Hakim}(1999)}]{brunel1999fast}%
  \BibitemOpen
  \bibfield  {author} {\bibinfo {author} {\bibfnamefont {N.}~\bibnamefont
  {Brunel}}\ and\ \bibinfo {author} {\bibfnamefont {V.}~\bibnamefont {Hakim}},\
  }\bibfield  {title} {\bibinfo {title} {Fast global oscillations in networks
  of integrate-and-fire neurons with low firing rates},\ }\href
  {https://direct.mit.edu/neco/article-abstract/11/7/1621/6299/Fast-Global-Oscillations-in-Networks-of-Integrate?redirectedFrom=fulltext}
  {\bibfield  {journal} {\bibinfo  {journal} {Neural Computation}\ }\textbf
  {\bibinfo {volume} {11}},\ \bibinfo {pages} {1621} (\bibinfo {year}
  {1999})}\BibitemShut {NoStop}%
\bibitem [{\citenamefont {Barzon}\ \emph {et~al.}(2022)\citenamefont {Barzon},
  \citenamefont {Nicoletti}, \citenamefont {Mariani}, \citenamefont
  {Formentin},\ and\ \citenamefont {Suweis}}]{barzon2022criticality}%
  \BibitemOpen
  \bibfield  {author} {\bibinfo {author} {\bibfnamefont {G.}~\bibnamefont
  {Barzon}}, \bibinfo {author} {\bibfnamefont {G.}~\bibnamefont {Nicoletti}},
  \bibinfo {author} {\bibfnamefont {B.}~\bibnamefont {Mariani}}, \bibinfo
  {author} {\bibfnamefont {M.}~\bibnamefont {Formentin}},\ and\ \bibinfo
  {author} {\bibfnamefont {S.}~\bibnamefont {Suweis}},\ }\bibfield  {title}
  {\bibinfo {title} {Criticality and network structure drive emergent
  oscillations in a stochastic whole-brain model},\ }\href@noop {} {\bibfield
  {journal} {\bibinfo  {journal} {Journal of Physics: Complexity}\ }\textbf
  {\bibinfo {volume} {3}},\ \bibinfo {pages} {025010} (\bibinfo {year}
  {2022})}\BibitemShut {NoStop}%
\bibitem [{\citenamefont {Huo}\ \emph {et~al.}(2022)\citenamefont {Huo},
  \citenamefont {Zou}, \citenamefont {Kaiser},\ and\ \citenamefont
  {Liu}}]{huo2022time}%
  \BibitemOpen
  \bibfield  {author} {\bibinfo {author} {\bibfnamefont {S.}~\bibnamefont
  {Huo}}, \bibinfo {author} {\bibfnamefont {Y.}~\bibnamefont {Zou}}, \bibinfo
  {author} {\bibfnamefont {M.}~\bibnamefont {Kaiser}},\ and\ \bibinfo {author}
  {\bibfnamefont {Z.}~\bibnamefont {Liu}},\ }\bibfield  {title} {\bibinfo
  {title} {Time-limited self-sustaining rhythms and state transitions in brain
  networks},\ }\href
  {https://journals.aps.org/prresearch/abstract/10.1103/PhysRevResearch.4.023076}
  {\bibfield  {journal} {\bibinfo  {journal} {Physical Review Research}\
  }\textbf {\bibinfo {volume} {4}},\ \bibinfo {pages} {023076} (\bibinfo {year}
  {2022})}\BibitemShut {NoStop}%
\bibitem [{\citenamefont {Griffis}\ \emph {et~al.}(2019)\citenamefont
  {Griffis}, \citenamefont {Metcalf}, \citenamefont {Corbetta},\ and\
  \citenamefont {Shulman}}]{griffis2019structural}%
  \BibitemOpen
  \bibfield  {author} {\bibinfo {author} {\bibfnamefont {J.~C.}\ \bibnamefont
  {Griffis}}, \bibinfo {author} {\bibfnamefont {N.~V.}\ \bibnamefont
  {Metcalf}}, \bibinfo {author} {\bibfnamefont {M.}~\bibnamefont {Corbetta}},\
  and\ \bibinfo {author} {\bibfnamefont {G.~L.}\ \bibnamefont {Shulman}},\
  }\bibfield  {title} {\bibinfo {title} {Structural disconnections explain
  brain network dysfunction after stroke},\ }\href
  {https://www.cell.com/cell-reports/fulltext/S2211-1247(19)31016-2} {\bibfield
   {journal} {\bibinfo  {journal} {Cell Reports}\ }\textbf {\bibinfo {volume}
  {28}},\ \bibinfo {pages} {2527} (\bibinfo {year} {2019})}\BibitemShut
  {NoStop}%
\bibitem [{\citenamefont {Rocha}\ \emph {et~al.}(2022)\citenamefont {Rocha},
  \citenamefont {Ko{\c{c}}illari}, \citenamefont {Suweis}, \citenamefont
  {De~Filippo De~Grazia}, \citenamefont {de~Schotten}, \citenamefont {Zorzi},\
  and\ \citenamefont {Corbetta}}]{rocha2022recovery}%
  \BibitemOpen
  \bibfield  {author} {\bibinfo {author} {\bibfnamefont {R.~P.}\ \bibnamefont
  {Rocha}}, \bibinfo {author} {\bibfnamefont {L.}~\bibnamefont
  {Ko{\c{c}}illari}}, \bibinfo {author} {\bibfnamefont {S.}~\bibnamefont
  {Suweis}}, \bibinfo {author} {\bibfnamefont {M.}~\bibnamefont {De~Filippo
  De~Grazia}}, \bibinfo {author} {\bibfnamefont {M.~T.}\ \bibnamefont
  {de~Schotten}}, \bibinfo {author} {\bibfnamefont {M.}~\bibnamefont {Zorzi}},\
  and\ \bibinfo {author} {\bibfnamefont {M.}~\bibnamefont {Corbetta}},\
  }\bibfield  {title} {\bibinfo {title} {Recovery of neural dynamics
  criticality in personalized whole-brain models of stroke},\ }\href
  {https://www.nature.com/articles/s41467-022-30892-6} {\bibfield  {journal}
  {\bibinfo  {journal} {Nature Communications}\ }\textbf {\bibinfo {volume}
  {13}},\ \bibinfo {pages} {3683} (\bibinfo {year} {2022})}\BibitemShut
  {NoStop}%
\bibitem [{\citenamefont {Rocha}\ \emph {et~al.}(2018)\citenamefont {Rocha},
  \citenamefont {Ko{\c{c}}illari}, \citenamefont {Suweis}, \citenamefont
  {Corbetta},\ and\ \citenamefont {Maritan}}]{rocha2018homeostatic}%
  \BibitemOpen
  \bibfield  {author} {\bibinfo {author} {\bibfnamefont {R.~P.}\ \bibnamefont
  {Rocha}}, \bibinfo {author} {\bibfnamefont {L.}~\bibnamefont
  {Ko{\c{c}}illari}}, \bibinfo {author} {\bibfnamefont {S.}~\bibnamefont
  {Suweis}}, \bibinfo {author} {\bibfnamefont {M.}~\bibnamefont {Corbetta}},\
  and\ \bibinfo {author} {\bibfnamefont {A.}~\bibnamefont {Maritan}},\
  }\bibfield  {title} {\bibinfo {title} {Homeostatic plasticity and emergence
  of functional networks in a whole-brain model at criticality},\ }\href
  {https://www.nature.com/articles/s41598-018-33923-9} {\bibfield  {journal}
  {\bibinfo  {journal} {Scientific Reports}\ }\textbf {\bibinfo {volume} {8}},\
  \bibinfo {pages} {15682} (\bibinfo {year} {2018})}\BibitemShut {NoStop}%
\bibitem [{\citenamefont {Jiang}\ \emph {et~al.}(2018)\citenamefont {Jiang},
  \citenamefont {Sui}, \citenamefont {Qiao}, \citenamefont {Dong},
  \citenamefont {Chen},\ and\ \citenamefont {Han}}]{jiang2018impaired}%
  \BibitemOpen
  \bibfield  {author} {\bibinfo {author} {\bibfnamefont {L.}~\bibnamefont
  {Jiang}}, \bibinfo {author} {\bibfnamefont {D.}~\bibnamefont {Sui}}, \bibinfo
  {author} {\bibfnamefont {K.}~\bibnamefont {Qiao}}, \bibinfo {author}
  {\bibfnamefont {H.-M.}\ \bibnamefont {Dong}}, \bibinfo {author}
  {\bibfnamefont {L.}~\bibnamefont {Chen}},\ and\ \bibinfo {author}
  {\bibfnamefont {Y.}~\bibnamefont {Han}},\ }\bibfield  {title} {\bibinfo
  {title} {Impaired functional criticality of human brain during alzheimer’s
  disease progression},\ }\href
  {https://www.nature.com/articles/s41598-018-19674-7} {\bibfield  {journal}
  {\bibinfo  {journal} {Scientific Reports}\ }\textbf {\bibinfo {volume} {8}},\
  \bibinfo {pages} {1324} (\bibinfo {year} {2018})}\BibitemShut {NoStop}%
\bibitem [{\citenamefont {Meisel}\ \emph {et~al.}(2012)\citenamefont {Meisel},
  \citenamefont {Storch}, \citenamefont {Hallmeyer-Elgner}, \citenamefont
  {Bullmore},\ and\ \citenamefont {Gross}}]{meisel2012failure}%
  \BibitemOpen
  \bibfield  {author} {\bibinfo {author} {\bibfnamefont {C.}~\bibnamefont
  {Meisel}}, \bibinfo {author} {\bibfnamefont {A.}~\bibnamefont {Storch}},
  \bibinfo {author} {\bibfnamefont {S.}~\bibnamefont {Hallmeyer-Elgner}},
  \bibinfo {author} {\bibfnamefont {E.}~\bibnamefont {Bullmore}},\ and\
  \bibinfo {author} {\bibfnamefont {T.}~\bibnamefont {Gross}},\ }\bibfield
  {title} {\bibinfo {title} {Failure of adaptive self-organized criticality
  during epileptic seizure attacks},\ }\href
  {https://journals.plos.org/ploscompbiol/article?id=10.1371/journal.pcbi.1002312}
  {\bibfield  {journal} {\bibinfo  {journal} {PLOS Computational Biology}\
  }\textbf {\bibinfo {volume} {8}},\ \bibinfo {pages} {e1002312} (\bibinfo
  {year} {2012})}\BibitemShut {NoStop}%
\bibitem [{\citenamefont {Meisel}(2020)}]{meisel2020antiepileptic}%
  \BibitemOpen
  \bibfield  {author} {\bibinfo {author} {\bibfnamefont {C.}~\bibnamefont
  {Meisel}},\ }\bibfield  {title} {\bibinfo {title} {Antiepileptic drugs induce
  subcritical dynamics in human cortical networks},\ }\href
  {https://www.pnas.org/doi/full/10.1073/pnas.1911461117} {\bibfield  {journal}
  {\bibinfo  {journal} {Proceedings of the National Academy of Sciences}\
  }\textbf {\bibinfo {volume} {117}},\ \bibinfo {pages} {11118} (\bibinfo
  {year} {2020})}\BibitemShut {NoStop}%
\bibitem [{\citenamefont {Siegel}\ \emph {et~al.}(2016)\citenamefont {Siegel},
  \citenamefont {Ramsey}, \citenamefont {Snyder}, \citenamefont {Metcalf},
  \citenamefont {Chacko}, \citenamefont {Weinberger}, \citenamefont
  {Baldassarre}, \citenamefont {Hacker}, \citenamefont {Shulman},\ and\
  \citenamefont {Corbetta}}]{siegel2016disruptions}%
  \BibitemOpen
  \bibfield  {author} {\bibinfo {author} {\bibfnamefont {J.~S.}\ \bibnamefont
  {Siegel}}, \bibinfo {author} {\bibfnamefont {L.~E.}\ \bibnamefont {Ramsey}},
  \bibinfo {author} {\bibfnamefont {A.~Z.}\ \bibnamefont {Snyder}}, \bibinfo
  {author} {\bibfnamefont {N.~V.}\ \bibnamefont {Metcalf}}, \bibinfo {author}
  {\bibfnamefont {R.~V.}\ \bibnamefont {Chacko}}, \bibinfo {author}
  {\bibfnamefont {K.}~\bibnamefont {Weinberger}}, \bibinfo {author}
  {\bibfnamefont {A.}~\bibnamefont {Baldassarre}}, \bibinfo {author}
  {\bibfnamefont {C.~D.}\ \bibnamefont {Hacker}}, \bibinfo {author}
  {\bibfnamefont {G.~L.}\ \bibnamefont {Shulman}},\ and\ \bibinfo {author}
  {\bibfnamefont {M.}~\bibnamefont {Corbetta}},\ }\bibfield  {title} {\bibinfo
  {title} {Disruptions of network connectivity predict impairment in multiple
  behavioral domains after stroke},\ }\href
  {https://www.pnas.org/doi/abs/10.1073/pnas.1521083113} {\bibfield  {journal}
  {\bibinfo  {journal} {Proceedings of the National Academy of Sciences}\
  }\textbf {\bibinfo {volume} {113}},\ \bibinfo {pages} {E4367} (\bibinfo
  {year} {2016})}\BibitemShut {NoStop}%
\bibitem [{\citenamefont {Corbetta}\ \emph {et~al.}(2015)\citenamefont
  {Corbetta}, \citenamefont {Ramsey}, \citenamefont {Callejas}, \citenamefont
  {Baldassarre}, \citenamefont {Hacker}, \citenamefont {Siegel}, \citenamefont
  {Astafiev}, \citenamefont {Rengachary}, \citenamefont {Zinn}, \citenamefont
  {Lang} \emph {et~al.}}]{corbetta2015common}%
  \BibitemOpen
  \bibfield  {author} {\bibinfo {author} {\bibfnamefont {M.}~\bibnamefont
  {Corbetta}}, \bibinfo {author} {\bibfnamefont {L.}~\bibnamefont {Ramsey}},
  \bibinfo {author} {\bibfnamefont {A.}~\bibnamefont {Callejas}}, \bibinfo
  {author} {\bibfnamefont {A.}~\bibnamefont {Baldassarre}}, \bibinfo {author}
  {\bibfnamefont {C.~D.}\ \bibnamefont {Hacker}}, \bibinfo {author}
  {\bibfnamefont {J.~S.}\ \bibnamefont {Siegel}}, \bibinfo {author}
  {\bibfnamefont {S.~V.}\ \bibnamefont {Astafiev}}, \bibinfo {author}
  {\bibfnamefont {J.}~\bibnamefont {Rengachary}}, \bibinfo {author}
  {\bibfnamefont {K.}~\bibnamefont {Zinn}}, \bibinfo {author} {\bibfnamefont
  {C.~E.}\ \bibnamefont {Lang}}, \emph {et~al.},\ }\bibfield  {title} {\bibinfo
  {title} {Common behavioral clusters and subcortical anatomy in stroke},\
  }\href {https://www.cell.com/neuron/fulltext/S0896-6273(15)00142-7}
  {\bibfield  {journal} {\bibinfo  {journal} {Neuron}\ }\textbf {\bibinfo
  {volume} {85}},\ \bibinfo {pages} {927} (\bibinfo {year} {2015})}\BibitemShut
  {NoStop}%
\bibitem [{\citenamefont {Ramsey}\ \emph {et~al.}(2017)\citenamefont {Ramsey},
  \citenamefont {Siegel}, \citenamefont {Lang}, \citenamefont {Strube},
  \citenamefont {Shulman},\ and\ \citenamefont
  {Corbetta}}]{ramsey2017behavioural}%
  \BibitemOpen
  \bibfield  {author} {\bibinfo {author} {\bibfnamefont {L.}~\bibnamefont
  {Ramsey}}, \bibinfo {author} {\bibfnamefont {J.}~\bibnamefont {Siegel}},
  \bibinfo {author} {\bibfnamefont {C.}~\bibnamefont {Lang}}, \bibinfo {author}
  {\bibfnamefont {M.}~\bibnamefont {Strube}}, \bibinfo {author} {\bibfnamefont
  {G.}~\bibnamefont {Shulman}},\ and\ \bibinfo {author} {\bibfnamefont
  {M.}~\bibnamefont {Corbetta}},\ }\bibfield  {title} {\bibinfo {title}
  {Behavioural clusters and predictors of performance during recovery from
  stroke},\ }\href {https://www.nature.com/articles/s41562-016-0038} {\bibfield
   {journal} {\bibinfo  {journal} {Nature Human Behaviour}\ }\textbf {\bibinfo
  {volume} {1}},\ \bibinfo {pages} {0038} (\bibinfo {year} {2017})}\BibitemShut
  {NoStop}%
\bibitem [{\citenamefont {Hagmann}\ \emph {et~al.}(2008)\citenamefont
  {Hagmann}, \citenamefont {Cammoun}, \citenamefont {Gigandet}, \citenamefont
  {Meuli}, \citenamefont {Honey}, \citenamefont {Wedeen},\ and\ \citenamefont
  {Sporns}}]{hagmann2008mapping}%
  \BibitemOpen
  \bibfield  {author} {\bibinfo {author} {\bibfnamefont {P.}~\bibnamefont
  {Hagmann}}, \bibinfo {author} {\bibfnamefont {L.}~\bibnamefont {Cammoun}},
  \bibinfo {author} {\bibfnamefont {X.}~\bibnamefont {Gigandet}}, \bibinfo
  {author} {\bibfnamefont {R.}~\bibnamefont {Meuli}}, \bibinfo {author}
  {\bibfnamefont {C.~J.}\ \bibnamefont {Honey}}, \bibinfo {author}
  {\bibfnamefont {V.~J.}\ \bibnamefont {Wedeen}},\ and\ \bibinfo {author}
  {\bibfnamefont {O.}~\bibnamefont {Sporns}},\ }\bibfield  {title} {\bibinfo
  {title} {Mapping the structural core of human cerebral cortex},\ }\href
  {https://journals.plos.org/plosbiology/article?id=10.1371/journal.pbio.0060159}
  {\bibfield  {journal} {\bibinfo  {journal} {PLOS Biology}\ }\textbf {\bibinfo
  {volume} {6}},\ \bibinfo {pages} {e159} (\bibinfo {year} {2008})}\BibitemShut
  {NoStop}%
\bibitem [{\citenamefont {Deco}\ \emph {et~al.}(2014)\citenamefont {Deco},
  \citenamefont {Ponce-Alvarez}, \citenamefont {Hagmann}, \citenamefont
  {Romani}, \citenamefont {Mantini},\ and\ \citenamefont
  {Corbetta}}]{deco2014local}%
  \BibitemOpen
  \bibfield  {author} {\bibinfo {author} {\bibfnamefont {G.}~\bibnamefont
  {Deco}}, \bibinfo {author} {\bibfnamefont {A.}~\bibnamefont {Ponce-Alvarez}},
  \bibinfo {author} {\bibfnamefont {P.}~\bibnamefont {Hagmann}}, \bibinfo
  {author} {\bibfnamefont {G.~L.}\ \bibnamefont {Romani}}, \bibinfo {author}
  {\bibfnamefont {D.}~\bibnamefont {Mantini}},\ and\ \bibinfo {author}
  {\bibfnamefont {M.}~\bibnamefont {Corbetta}},\ }\bibfield  {title} {\bibinfo
  {title} {How local excitation--inhibition ratio impacts the whole brain
  dynamics},\ }\href@noop {} {\bibfield  {journal} {\bibinfo  {journal}
  {Journal of Neuroscience}\ }\textbf {\bibinfo {volume} {34}},\ \bibinfo
  {pages} {7886} (\bibinfo {year} {2014})}\BibitemShut {NoStop}%
\bibitem [{\citenamefont {Hellyer}\ \emph {et~al.}(2016)\citenamefont
  {Hellyer}, \citenamefont {Jachs}, \citenamefont {Clopath},\ and\
  \citenamefont {Leech}}]{hellyer2016local}%
  \BibitemOpen
  \bibfield  {author} {\bibinfo {author} {\bibfnamefont {P.~J.}\ \bibnamefont
  {Hellyer}}, \bibinfo {author} {\bibfnamefont {B.}~\bibnamefont {Jachs}},
  \bibinfo {author} {\bibfnamefont {C.}~\bibnamefont {Clopath}},\ and\ \bibinfo
  {author} {\bibfnamefont {R.}~\bibnamefont {Leech}},\ }\bibfield  {title}
  {\bibinfo {title} {Local inhibitory plasticity tunes macroscopic brain
  dynamics and allows the emergence of functional brain networks},\ }\href@noop
  {} {\bibfield  {journal} {\bibinfo  {journal} {NeuroImage}\ }\textbf
  {\bibinfo {volume} {124}},\ \bibinfo {pages} {85} (\bibinfo {year}
  {2016})}\BibitemShut {NoStop}%
\bibitem [{\citenamefont {Abeysuriya}\ \emph {et~al.}(2018)\citenamefont
  {Abeysuriya}, \citenamefont {Hadida}, \citenamefont {Sotiropoulos},
  \citenamefont {Jbabdi}, \citenamefont {Becker}, \citenamefont {Hunt},
  \citenamefont {Brookes},\ and\ \citenamefont
  {Woolrich}}]{abeysuriya2018biophysical}%
  \BibitemOpen
  \bibfield  {author} {\bibinfo {author} {\bibfnamefont {R.~G.}\ \bibnamefont
  {Abeysuriya}}, \bibinfo {author} {\bibfnamefont {J.}~\bibnamefont {Hadida}},
  \bibinfo {author} {\bibfnamefont {S.~N.}\ \bibnamefont {Sotiropoulos}},
  \bibinfo {author} {\bibfnamefont {S.}~\bibnamefont {Jbabdi}}, \bibinfo
  {author} {\bibfnamefont {R.}~\bibnamefont {Becker}}, \bibinfo {author}
  {\bibfnamefont {B.~A.}\ \bibnamefont {Hunt}}, \bibinfo {author}
  {\bibfnamefont {M.~J.}\ \bibnamefont {Brookes}},\ and\ \bibinfo {author}
  {\bibfnamefont {M.~W.}\ \bibnamefont {Woolrich}},\ }\bibfield  {title}
  {\bibinfo {title} {A biophysical model of dynamic balancing of excitation and
  inhibition in fast oscillatory large-scale networks},\ }\href@noop {}
  {\bibfield  {journal} {\bibinfo  {journal} {PLOS Computational Biology}\
  }\textbf {\bibinfo {volume} {14}},\ \bibinfo {pages} {e1006007} (\bibinfo
  {year} {2018})}\BibitemShut {NoStop}%
\bibitem [{\citenamefont {Gordon}\ \emph {et~al.}(2016)\citenamefont {Gordon},
  \citenamefont {Laumann}, \citenamefont {Adeyemo}, \citenamefont {Huckins},
  \citenamefont {Kelley},\ and\ \citenamefont
  {Petersen}}]{gordon2016generation}%
  \BibitemOpen
  \bibfield  {author} {\bibinfo {author} {\bibfnamefont {E.~M.}\ \bibnamefont
  {Gordon}}, \bibinfo {author} {\bibfnamefont {T.~O.}\ \bibnamefont {Laumann}},
  \bibinfo {author} {\bibfnamefont {B.}~\bibnamefont {Adeyemo}}, \bibinfo
  {author} {\bibfnamefont {J.~F.}\ \bibnamefont {Huckins}}, \bibinfo {author}
  {\bibfnamefont {W.~M.}\ \bibnamefont {Kelley}},\ and\ \bibinfo {author}
  {\bibfnamefont {S.~E.}\ \bibnamefont {Petersen}},\ }\bibfield  {title}
  {\bibinfo {title} {Generation and evaluation of a cortical area parcellation
  from resting-state correlations},\ }\href
  {https://academic.oup.com/cercor/article/26/1/288/2367115} {\bibfield
  {journal} {\bibinfo  {journal} {Cerebral Cortex}\ }\textbf {\bibinfo {volume}
  {26}},\ \bibinfo {pages} {288} (\bibinfo {year} {2016})}\BibitemShut
  {NoStop}%
\bibitem [{\citenamefont {Carmichael}(2016)}]{carmichael20163}%
  \BibitemOpen
  \bibfield  {author} {\bibinfo {author} {\bibfnamefont {S.~T.}\ \bibnamefont
  {Carmichael}},\ }\bibfield  {title} {\bibinfo {title} {The 3 rs of stroke
  biology: radial, relayed, and regenerative},\ }\href@noop {} {\bibfield
  {journal} {\bibinfo  {journal} {Neurotherapeutics}\ }\textbf {\bibinfo
  {volume} {13}},\ \bibinfo {pages} {348} (\bibinfo {year} {2016})}\BibitemShut
  {NoStop}%
\bibitem [{\citenamefont {Carmichael}(2010)}]{carmichael2010targets}%
  \BibitemOpen
  \bibfield  {author} {\bibinfo {author} {\bibfnamefont {S.~T.}\ \bibnamefont
  {Carmichael}},\ }\bibfield  {title} {\bibinfo {title} {Targets for neural
  repair therapies after stroke},\ }\href@noop {} {\bibfield  {journal}
  {\bibinfo  {journal} {Stroke}\ }\textbf {\bibinfo {volume} {41}},\ \bibinfo
  {pages} {S124} (\bibinfo {year} {2010})}\BibitemShut {NoStop}%
\bibitem [{\citenamefont {Shi}\ \emph {et~al.}(2019)\citenamefont {Shi},
  \citenamefont {Tian}, \citenamefont {Li}, \citenamefont {Ducruet},
  \citenamefont {Lawton},\ and\ \citenamefont {Shi}}]{shi2019global}%
  \BibitemOpen
  \bibfield  {author} {\bibinfo {author} {\bibfnamefont {K.}~\bibnamefont
  {Shi}}, \bibinfo {author} {\bibfnamefont {D.-C.}\ \bibnamefont {Tian}},
  \bibinfo {author} {\bibfnamefont {Z.-G.}\ \bibnamefont {Li}}, \bibinfo
  {author} {\bibfnamefont {A.~F.}\ \bibnamefont {Ducruet}}, \bibinfo {author}
  {\bibfnamefont {M.~T.}\ \bibnamefont {Lawton}},\ and\ \bibinfo {author}
  {\bibfnamefont {F.-D.}\ \bibnamefont {Shi}},\ }\bibfield  {title} {\bibinfo
  {title} {Global brain inflammation in stroke},\ }\href@noop {} {\bibfield
  {journal} {\bibinfo  {journal} {The Lancet Neurology}\ }\textbf {\bibinfo
  {volume} {18}},\ \bibinfo {pages} {1058} (\bibinfo {year}
  {2019})}\BibitemShut {NoStop}%
\bibitem [{\citenamefont {Carter}\ \emph {et~al.}(2012)\citenamefont {Carter},
  \citenamefont {Shulman},\ and\ \citenamefont {Corbetta}}]{carter2012use}%
  \BibitemOpen
  \bibfield  {author} {\bibinfo {author} {\bibfnamefont {A.~R.}\ \bibnamefont
  {Carter}}, \bibinfo {author} {\bibfnamefont {G.~L.}\ \bibnamefont
  {Shulman}},\ and\ \bibinfo {author} {\bibfnamefont {M.}~\bibnamefont
  {Corbetta}},\ }\bibfield  {title} {\bibinfo {title} {Why use a
  connectivity-based approach to study stroke and recovery of function?},\
  }\href@noop {} {\bibfield  {journal} {\bibinfo  {journal} {NeuroImage}\
  }\textbf {\bibinfo {volume} {62}},\ \bibinfo {pages} {2271} (\bibinfo {year}
  {2012})}\BibitemShut {NoStop}%
\bibitem [{\citenamefont {Guggisberg}\ \emph {et~al.}(2019)\citenamefont
  {Guggisberg}, \citenamefont {Koch}, \citenamefont {Hummel},\ and\
  \citenamefont {Buetefisch}}]{guggisberg2019brain}%
  \BibitemOpen
  \bibfield  {author} {\bibinfo {author} {\bibfnamefont {A.~G.}\ \bibnamefont
  {Guggisberg}}, \bibinfo {author} {\bibfnamefont {P.~J.}\ \bibnamefont
  {Koch}}, \bibinfo {author} {\bibfnamefont {F.~C.}\ \bibnamefont {Hummel}},\
  and\ \bibinfo {author} {\bibfnamefont {C.~M.}\ \bibnamefont {Buetefisch}},\
  }\bibfield  {title} {\bibinfo {title} {Brain networks and their relevance for
  stroke rehabilitation},\ }\href@noop {} {\bibfield  {journal} {\bibinfo
  {journal} {Clinical Neurophysiology}\ }\textbf {\bibinfo {volume} {130}},\
  \bibinfo {pages} {1098} (\bibinfo {year} {2019})}\BibitemShut {NoStop}%
\bibitem [{\citenamefont {Siegel}\ \emph {et~al.}(2018)\citenamefont {Siegel},
  \citenamefont {Seitzman}, \citenamefont {Ramsey}, \citenamefont {Ortega},
  \citenamefont {Gordon}, \citenamefont {Dosenbach}, \citenamefont {Petersen},
  \citenamefont {Shulman},\ and\ \citenamefont {Corbetta}}]{siegel2018re}%
  \BibitemOpen
  \bibfield  {author} {\bibinfo {author} {\bibfnamefont {J.~S.}\ \bibnamefont
  {Siegel}}, \bibinfo {author} {\bibfnamefont {B.~A.}\ \bibnamefont
  {Seitzman}}, \bibinfo {author} {\bibfnamefont {L.~E.}\ \bibnamefont
  {Ramsey}}, \bibinfo {author} {\bibfnamefont {M.}~\bibnamefont {Ortega}},
  \bibinfo {author} {\bibfnamefont {E.~M.}\ \bibnamefont {Gordon}}, \bibinfo
  {author} {\bibfnamefont {N.~U.}\ \bibnamefont {Dosenbach}}, \bibinfo {author}
  {\bibfnamefont {S.~E.}\ \bibnamefont {Petersen}}, \bibinfo {author}
  {\bibfnamefont {G.~L.}\ \bibnamefont {Shulman}},\ and\ \bibinfo {author}
  {\bibfnamefont {M.}~\bibnamefont {Corbetta}},\ }\bibfield  {title} {\bibinfo
  {title} {Re-emergence of modular brain networks in stroke recovery},\
  }\href@noop {} {\bibfield  {journal} {\bibinfo  {journal} {Cortex}\ }\textbf
  {\bibinfo {volume} {101}},\ \bibinfo {pages} {44} (\bibinfo {year}
  {2018})}\BibitemShut {NoStop}%
\bibitem [{\citenamefont {Gratton}\ \emph {et~al.}(2012)\citenamefont
  {Gratton}, \citenamefont {Nomura}, \citenamefont {P{\'e}rez},\ and\
  \citenamefont {D'Esposito}}]{gratton2012focal}%
  \BibitemOpen
  \bibfield  {author} {\bibinfo {author} {\bibfnamefont {C.}~\bibnamefont
  {Gratton}}, \bibinfo {author} {\bibfnamefont {E.~M.}\ \bibnamefont {Nomura}},
  \bibinfo {author} {\bibfnamefont {F.}~\bibnamefont {P{\'e}rez}},\ and\
  \bibinfo {author} {\bibfnamefont {M.}~\bibnamefont {D'Esposito}},\ }\bibfield
   {title} {\bibinfo {title} {Focal brain lesions to critical locations cause
  widespread disruption of the modular organization of the brain},\ }\href@noop
  {} {\bibfield  {journal} {\bibinfo  {journal} {Journal of Cognitive
  Neuroscience}\ }\textbf {\bibinfo {volume} {24}},\ \bibinfo {pages} {1275}
  (\bibinfo {year} {2012})}\BibitemShut {NoStop}%
\bibitem [{\citenamefont {Blondel}\ \emph {et~al.}(2008)\citenamefont
  {Blondel}, \citenamefont {Guillaume}, \citenamefont {Lambiotte},\ and\
  \citenamefont {Lefebvre}}]{Blondel2008}%
  \BibitemOpen
  \bibfield  {author} {\bibinfo {author} {\bibfnamefont {V.~D.}\ \bibnamefont
  {Blondel}}, \bibinfo {author} {\bibfnamefont {J.~L.}\ \bibnamefont
  {Guillaume}}, \bibinfo {author} {\bibfnamefont {R.}~\bibnamefont
  {Lambiotte}},\ and\ \bibinfo {author} {\bibfnamefont {E.}~\bibnamefont
  {Lefebvre}},\ }\bibfield  {title} {\bibinfo {title} {Fast unfolding of
  communities in large networks},\ }\href@noop {} {\bibfield  {journal}
  {\bibinfo  {journal} {Journal of Statistical Mechanics: Theory and
  Experiment}\ }\textbf {\bibinfo {volume} {2008}},\ \bibinfo {pages} {P10008}
  (\bibinfo {year} {2008})}\BibitemShut {NoStop}%
\bibitem [{\citenamefont {Lakens}(2013)}]{lakens2013calculating}%
  \BibitemOpen
  \bibfield  {author} {\bibinfo {author} {\bibfnamefont {D.}~\bibnamefont
  {Lakens}},\ }\bibfield  {title} {\bibinfo {title} {Calculating and reporting
  effect sizes to facilitate cumulative science: a practical primer for t-tests
  and anovas},\ }\href@noop {} {\bibfield  {journal} {\bibinfo  {journal}
  {Frontiers in Psychology}\ }\textbf {\bibinfo {volume} {4}},\ \bibinfo
  {pages} {863} (\bibinfo {year} {2013})}\BibitemShut {NoStop}%
\bibitem [{\citenamefont {Haimovici}\ \emph {et~al.}(2013)\citenamefont
  {Haimovici}, \citenamefont {Tagliazucchi}, \citenamefont {Balenzuela},\ and\
  \citenamefont {Chialvo}}]{haimovici2013brain}%
  \BibitemOpen
  \bibfield  {author} {\bibinfo {author} {\bibfnamefont {A.}~\bibnamefont
  {Haimovici}}, \bibinfo {author} {\bibfnamefont {E.}~\bibnamefont
  {Tagliazucchi}}, \bibinfo {author} {\bibfnamefont {P.}~\bibnamefont
  {Balenzuela}},\ and\ \bibinfo {author} {\bibfnamefont {D.~R.}\ \bibnamefont
  {Chialvo}},\ }\bibfield  {title} {\bibinfo {title} {Brain organization into
  resting state networks emerges at criticality on a model of the human
  connectome},\ }\href
  {https://journals.aps.org/prl/abstract/10.1103/PhysRevLett.110.178101}
  {\bibfield  {journal} {\bibinfo  {journal} {Physical Review Letters}\
  }\textbf {\bibinfo {volume} {110}},\ \bibinfo {pages} {178101} (\bibinfo
  {year} {2013})}\BibitemShut {NoStop}%
\bibitem [{\citenamefont {Crepey}\ \emph {et~al.}(2006)\citenamefont {Crepey},
  \citenamefont {Alvarez},\ and\ \citenamefont
  {Barth{\'e}lemy}}]{crepey2006epidemic}%
  \BibitemOpen
  \bibfield  {author} {\bibinfo {author} {\bibfnamefont {P.}~\bibnamefont
  {Crepey}}, \bibinfo {author} {\bibfnamefont {F.~P.}\ \bibnamefont
  {Alvarez}},\ and\ \bibinfo {author} {\bibfnamefont {M.}~\bibnamefont
  {Barth{\'e}lemy}},\ }\bibfield  {title} {\bibinfo {title} {Epidemic
  variability in complex networks},\ }\href
  {https://journals.aps.org/pre/abstract/10.1103/PhysRevE.73.046131} {\bibfield
   {journal} {\bibinfo  {journal} {Physical Review E}\ }\textbf {\bibinfo
  {volume} {73}},\ \bibinfo {pages} {046131} (\bibinfo {year}
  {2006})}\BibitemShut {NoStop}%
\bibitem [{\citenamefont {Yang}\ \emph {et~al.}(2023)\citenamefont {Yang},
  \citenamefont {Wu}, \citenamefont {He}, \citenamefont {Xu},\ and\
  \citenamefont {Zheng}}]{yang2023asymmetric}%
  \BibitemOpen
  \bibfield  {author} {\bibinfo {author} {\bibfnamefont {Z.}~\bibnamefont
  {Yang}}, \bibinfo {author} {\bibfnamefont {J.}~\bibnamefont {Wu}}, \bibinfo
  {author} {\bibfnamefont {J.}~\bibnamefont {He}}, \bibinfo {author}
  {\bibfnamefont {K.}~\bibnamefont {Xu}},\ and\ \bibinfo {author}
  {\bibfnamefont {M.}~\bibnamefont {Zheng}},\ }\bibfield  {title} {\bibinfo
  {title} {Asymmetric inter-layer interactions induce a double transition of
  information spreading},\ }\href@noop {} {\bibfield  {journal} {\bibinfo
  {journal} {Chaos, Solitons \& Fractals}\ }\textbf {\bibinfo {volume} {171}},\
  \bibinfo {pages} {113487} (\bibinfo {year} {2023})}\BibitemShut {NoStop}%
\bibitem [{\citenamefont {Janarek}\ \emph {et~al.}(2023)\citenamefont
  {Janarek}, \citenamefont {Drogosz}, \citenamefont {Grela}, \citenamefont
  {Ochab},\ and\ \citenamefont
  {O{\'s}wi{\k{e}}cimka}}]{janarek2023investigating}%
  \BibitemOpen
  \bibfield  {author} {\bibinfo {author} {\bibfnamefont {J.}~\bibnamefont
  {Janarek}}, \bibinfo {author} {\bibfnamefont {Z.}~\bibnamefont {Drogosz}},
  \bibinfo {author} {\bibfnamefont {J.}~\bibnamefont {Grela}}, \bibinfo
  {author} {\bibfnamefont {J.~K.}\ \bibnamefont {Ochab}},\ and\ \bibinfo
  {author} {\bibfnamefont {P.}~\bibnamefont {O{\'s}wi{\k{e}}cimka}},\
  }\bibfield  {title} {\bibinfo {title} {Investigating structural and
  functional aspects of the brain’s criticality in stroke},\ }\href@noop {}
  {\bibfield  {journal} {\bibinfo  {journal} {Scientific Reports}\ }\textbf
  {\bibinfo {volume} {13}},\ \bibinfo {pages} {12341} (\bibinfo {year}
  {2023})}\BibitemShut {NoStop}%
\bibitem [{\citenamefont {Breakspear}(2017)}]{breakspear2017dynamic}%
  \BibitemOpen
  \bibfield  {author} {\bibinfo {author} {\bibfnamefont {M.}~\bibnamefont
  {Breakspear}},\ }\bibfield  {title} {\bibinfo {title} {Dynamic models of
  large-scale brain activity},\ }\href@noop {} {\bibfield  {journal} {\bibinfo
  {journal} {Nature Neuroscience}\ }\textbf {\bibinfo {volume} {20}},\ \bibinfo
  {pages} {340} (\bibinfo {year} {2017})}\BibitemShut {NoStop}%
\bibitem [{\citenamefont {Qian}\ \emph {et~al.}(2010)\citenamefont {Qian},
  \citenamefont {Huang}, \citenamefont {Hu},\ and\ \citenamefont
  {Liao}}]{qian2010structure}%
  \BibitemOpen
  \bibfield  {author} {\bibinfo {author} {\bibfnamefont {Y.}~\bibnamefont
  {Qian}}, \bibinfo {author} {\bibfnamefont {X.}~\bibnamefont {Huang}},
  \bibinfo {author} {\bibfnamefont {G.}~\bibnamefont {Hu}},\ and\ \bibinfo
  {author} {\bibfnamefont {X.}~\bibnamefont {Liao}},\ }\bibfield  {title}
  {\bibinfo {title} {Structure and control of self-sustained target waves in
  excitable small-world networks},\ }\href@noop {} {\bibfield  {journal}
  {\bibinfo  {journal} {Physical Review E}\ }\textbf {\bibinfo {volume} {81}},\
  \bibinfo {pages} {036101} (\bibinfo {year} {2010})}\BibitemShut {NoStop}%
\bibitem [{\citenamefont {Liao}\ \emph {et~al.}(2011)\citenamefont {Liao},
  \citenamefont {Qian}, \citenamefont {Mi}, \citenamefont {Xia}, \citenamefont
  {Huang},\ and\ \citenamefont {Hu}}]{liao2011oscillation}%
  \BibitemOpen
  \bibfield  {author} {\bibinfo {author} {\bibfnamefont {X.-H.}\ \bibnamefont
  {Liao}}, \bibinfo {author} {\bibfnamefont {Y.}~\bibnamefont {Qian}}, \bibinfo
  {author} {\bibfnamefont {Y.-Y.}\ \bibnamefont {Mi}}, \bibinfo {author}
  {\bibfnamefont {Q.-Z.}\ \bibnamefont {Xia}}, \bibinfo {author} {\bibfnamefont
  {X.-Q.}\ \bibnamefont {Huang}},\ and\ \bibinfo {author} {\bibfnamefont
  {G.}~\bibnamefont {Hu}},\ }\bibfield  {title} {\bibinfo {title} {Oscillation
  sources and wave propagation paths in complex networks consisting of
  excitable nodes},\ }\href@noop {} {\bibfield  {journal} {\bibinfo  {journal}
  {Frontiers of Physics in China}\ }\textbf {\bibinfo {volume} {6}},\ \bibinfo
  {pages} {124} (\bibinfo {year} {2011})}\BibitemShut {NoStop}%
\end{thebibliography}%
%

\end{document}